\documentclass[mathleft
]{an}
\usepackage{graphicx}
\usepackage{times}

\usepackage{amsmath}
\usepackage{amssymb}

\usepackage{color}

\newcommand{\hst}{{\sl HST}}

\begin{document}

\Pagespan{001}{}
\Yearpublication{2013}%
\Yearsubmission{2013}%
\Month{00}%
\Volume{000}%
\Issue{00}%
\DOI{This.is/not.aDOI}%

\title{
The M\,4 Core Project with \textit{HST\/} -- I. ~ Overview and First-Epoch\thanks{
%
Based on observations collected with the NASA/ESA Hubble Space
Telescope, obtained at the Space Telescope Science Institute, which is
operated by AURA, Inc., under NASA contract NAS 5-26555, under large program 
GO-12911 .
}}

\author{L.\ R.\ Bedin\inst{1}\fnmsep\thanks{Corresponding author:
    \email{luigi.bedin@inaf.oapd.it}\newline},
J.\ Anderson\inst{2}, 
D.\ C.\ Heggie\inst{3}, 
G.\ Piotto\inst{4,1}, 
A.\ P.\ Milone\inst{5,6,7}, 
M.\ Giersz\inst{8}, 
V.~Nascimbeni\inst{4,1}, 
A.\ Bellini\inst{2}, 
R.\ M. Rich\inst{9},   
M.\ van den Berg\inst{10,11}, 
D.\ Pooley\inst{12,13}, 
K.\ Brogaard\inst{14,15}, 
S.~Ortolani\inst{4,1},  
L.\ Malavolta\inst{4,1}, 
L.\ Ubeda\inst{2}, 
\and  
A.\ F.\ Marino\inst{7} 
}
\titlerunning{The M\,4 Core Project with \hst}
\authorrunning{L.\ R.\ Bedin et al.}
\institute{
Istituto Nazionale di Astrofisica - Osservatorio Astronomico di Padova, 
Vicolo dell'Osservatorio 5, Padova, IT-35122
\and
Space Telescope Science Institute, 3800 San Martin Drive, Baltimore, MD 21218, USA
\and 
School of Mathematics and Maxwell Institute for Mathematical Sciences, 
University of Edinburgh, Kings Buildings, Edinburgh, UK-EH9-3JZ
\and 
Dipartimento di Fisica e Astronomia ``Galileo Galilei'', Univ.
di Padova, Vicolo dell'Osservatorio 3, Padova IT-35122
\and 
Instituto de Astrof\`isica de Canarias, La Laguna, Tenerife, Canary Islands, ES-38200
\and 
Department of Astrophysics, University of La Laguna, La Laguna, Tenerife, Canary Islands, ES-38200
\and 
Research School of Astronomy and Astrophysics, The Australian National University, 
Cotter Road, Weston, ACT, 2611, Australia
\and 
Nicolaus Copernicus Astronomical Center, Polish Academy of Sciences, ul. 
Bartycka 18, 00-716, Warsaw, Poland
\and 
Department of Physics and Astronomy, University of California, Los Angeles, CA 90095, USA
\and 
Astronomical Institute ``Anton Pannekoek'', University of Amsterdam, Science Park 904, 
1098 XH Amsterdam, The Netherlands
\and 
Harvard-Smithsonian Center for Astrophysics, 60 Garden Street, Cambridge, 02138 MA, USA
\and 
Department of Physics, Sam Houston State University, Huntsville, TX 77341, USA
\and
Eureka Scientific, Inc., 2452 Delmer Street, Suite 100, Oakland, CA 94602, USA
\and 
Department of Physics and Astronomy, Aarhus University, Ny Munkegade, 8000 Aarhus C, Denmark
\and 
Department of Physics and Astronomy, University of Victoria, PO Box 3055, Victoria, B.C., 
V8W 3P6, Canada
}

\received{ 12 February 2013}
\accepted{27 May 2013}
\publonline{later}


\abstract{
We present an overview of the ongoing \textit{Hubble Space Telescope}
large program GO-12911.  The program is focused on the core of M\,4,
the nearest Galactic globular cluster, and the observations are
designed to constrain the number of binaries with massive companions
(black holes, neutron stars, or white dwarfs) by measuring the
``wobble'' of the luminous (main-sequence) companion around the center
of mass of the pair, with an astrometric precision of $\sim$50
$\mu$as.
The high spatial resolution and stable medium-band PSFs of WFC3/UVIS will 
make these measurements possible.
In this work we describe: 
    \textit{(i)\/}   the motivation behind this study, 
    \textit{(ii)\/}  our observing strategy, 
    \textit{(iii)\/} the many other investigations 
                     enabled by this \textit{unique} data set, 
                     and which of those our team is conducting, and  
    \textit{(iv)\/}  a preliminary reduction of the first-epoch 
                     data-set collected on October 10, 2012.
}

\maketitle

\section{Introduction}
\label{intro}
The \textit{Hubble Space Telescope\/} (\textit{HST\/}\,) Large Program
entitled: \textit{``A search for binaries with massive companions in
  the core of the closest globular cluster M\,4''} (program GO-12911,
PI: Bedin, for which 120 orbits have been allocated during
\textit{HST\/}'s Cycle\,20), has begun to collect data.
The project takes advantage of the exquisitely high spatial resolution
and astrometric capabilities of \textit{HST\/} to measure the
``wobble'' 
(i.e., back and forth motions, at the 50 micro-arcsec level)
of luminous companions orbiting around massive dark-remnant
primaries in the core of a Galactic globular cluster (GC), where
massive binaries are expected to have sunk because of mass
segregation.
The target GC is M\,4 (NGC\,6121), the cluster geometrically closest to
us and known to be rich in binaries.
This project also has a spectroscopic counter-part involving about
10\,000 FLAMES@VLT spectra aimed at searching for and characterizing
the binary population in M\,4, by measuring radial-velocity
variations, mainly outside the core region (Sommariva et al.\ 2009 and
Malavolta et al.\ 2013, in preparation).

This is the first paper of a series dedicated to the survey, and its
aim is to provide the astronomical community with an overview of the
program.
In Sect.~\ref{mot} we summarize the motivation behind this project,
while the choice of the target is explained in Sect.~\ref{tgt}.
Descriptions of the expected properties of the binaries, and how to
detect their wobbles are given in Sect.~\ref{expected} and
\ref{detection}, respectively.  Section~\ref{obsstg} outlines in great
detail the observing strategy, and how we ended up with the current
plan.
Section~\ref{cmp} compares this new survey with the previous large
surveys of a GC carried out by \hst.
To illustrate the typical data we will have for each of the 12 epochs,
we present in Sect.~\ref{1ste} the preliminary reduction of the
primary- and parallel-observations collected during the first epoch of
observations.
We conclude this work in Sect.\ 9, with a description of astrophysical
(and technical) studies which our team is conducting.

\section{Motivation}
\label{mot}

The frequency of binary stars is a fundamental property of a stellar
population, and this is particularly true for globular clusters, where
stars are close enough to interact frequently.  Binaries have a much
larger interaction cross section than single stars, and if they are
present they catalyze the disruption of widely separated binaries and
the formation of exotic objects, such as blue stragglers, cataclysmic
variables, and millisecond pulsars.  Ultimately, binaries regulate the
long-term evolution of the cluster.

In globular clusters, interactions between binary stars play the same
role as nuclear reactions in stars:\ they prevent collapse of the
core.  For many years, searches for binaries in clusters revealed only
a very small binary fraction, which led theorists to conclude that
there were almost no primordial binaries (or none left) and that
binaries would only be formed in the high densities reached at the end
of core collapse, heralding a succession of core oscillations as the
cluster steadily lost stars across the Galactic tidal boundary.
Recently, as we have been able to search for binaries well into the
core using multi-wavelength photometric techniques, it is clear that
they are quite numerous (see Milone et al.\ 2012a, and references
therein), and we had previously missed them because they had sunk to
the core.
The presence of primordial binaries in clusters has a significant
effect on our dynamical models.  Theoretical modeling has shown that
the massive primordial binaries will segregate to the core and come to
dominate the dynamical
interactions there.  The hardening
of binaries in the core halts core collapse at much lower central
densities, and leads to a long phase of near-equilibrium in the core,
while stars in the outer envelope of the cluster slowly evaporate over
the tidal boundary (see Meylan \& Heggie 1997, or Heggie \& Hut 2003 
for reviews).

This picture raises an interesting paradox: if binaries can be shown
to prevent deep core collapse, why do we observe any
post-core-collapse clusters at all?\footnote{
Note that, what observers call ``post-core-collapse''
clusters are objects which cannot be fitted by a King model, and the
remainder are often referred to as ``non-core-collapse'' clusters.
Clearly the names given to these two types of cluster imply a
particular theoretical interpretation of the shape of the surface
brightness profile: it is assumed that clusters with a King profile
are still in the process of core collapse.  Moreover, it is often
assumed that the post-core-collapse clusters are objects in a steady
binary burning phase.  These interpretations are, however,
controversial.
}
For example, M\,4, our target, shows no evidence for any central
brightness cusp (Trager et al.\ 1995), despite the fact that its age
($\sim$12 Gyr, Bedin et al.\ 2009) exceeds its present central
relaxation time ($\sim$0.08 Gyr, Harris 1996, as updated in 2010) by a
factor $\sim$150.  Can the lack of evidence of core collapse be traced
to the presence of a large fraction of binaries?  
Let us furthermore compare
M\,4 with NGC\,6397. NGC\,6397 has a mass, relaxation time, and orbit
similar to those of M\,4.  Therefore it should be at a similar stage
of its evolution, and yet it has a completely different structure with
a collapsed core.  Is this due to a relative lack of primordial
binaries?  (Compare the estimated overall fraction of binaries by Milone et
al.\ 2012a: 10\% in the core of M\,4 and $\sim$2\% for
NGC\,6397).  There is another possible explanation for the difference
between these two clusters, however.  Even in a core sustained by
primordial binaries, large oscillations in the core density are
possible, and it may be that such clusters oscillate between collapsed
and uncollapsed structures (Heggie \& Giersz 2009).

Up to now it has been impossible to make these ideas quantitative,
because the properties of the binary population are almost completely
unconstrained by observations.  Modelling the dynamical evolution with
binaries, on the other hand, is straightforward.  It is true that the
largest $N$-body model of this kind published so far starts with only
$2.5\times10^5$ stars (Sippel \& Hurley 2013), about half the
estimated original number of stars in M\,4 (Heggie \& Giersz 2008),
and even one such model takes months of computing.  Until software
suitable for clusters of GPUs becomes available, simpler methods than
$N$-body methods are used, especially the Monte Carlo method.  With
this method a full-sized model of M\,4 takes just days, and extensive
testing against $N$-body models (in the range of $N$ reachable by
$N$-body codes) shows very good agreement (Giersz, Heggie \& Hurley
2008; Giersz et al.\ 2013).

Some years ago some of us constructed a Monte Carlo dynamical
evolutionary model for M\,4 (Heggie \& Giersz, 2008). It is almost as
elaborate as an $N$-body model, and includes appropriate stellar
initial mass functions, a primordial binary population, Galactic tidal
effects, synthetic stellar and binary evolution, relaxation, and
three- and four-body interactions.  By adjustment of the initial
conditions, the model was brought into satisfactory agreement with the
present-day surface brightness and velocity dispersion profiles, and
the luminosity function (LF).  This was the first time such a
comprehensive, realistic evolutionary model had been constructed for
any globular cluster. One conclusion from this model was that M\,4 was
indeed found to be a post-collapse cluster sustained by binary
burning.  As a by-product, the model gave detailed predictions for the
distribution of the binaries in space and period.

Such a comparison between a model and the observed parameters is of
fundamental importance to understanding the dynamical evolution of
GCs, rather like what happens when we compare the observed
color-magnitude diagrams and luminosity functions with stellar
evolution models.
Understanding the dynamical evolution of GCs also allows us to better 
understand their origin (and therefore the origin of the Galaxy), and the 
evolution and properties of their stellar populations (including exotica). 
Close interaction between observers and theorists is vital to progress
here, and our program is a key project within an international
collaboration, MODEST (Modeling of Dense Stellar Systems, {\sf
  http://www.manybody.org/modest}), within which experts in stellar
dynamics, stellar evolution, and observers have joined their efforts
for a comprehensive study of the origin and evolution of star
clusters.

The main limitation in the modelling of M\,4 was our almost complete
ignorance of the properties of the binaries, except for the binary
fraction.
Our aim in this program is to provide the observational constraints on
the binary population which the modelling needs, and so to test our
theoretical picture of M\,4 against observations for the first
time. The present investigation, coupled with the radial velocity data
already collected at VLT/FLAMES (the $\sim$ 6000 spectra discussed in
Sommariva et al.\ 2009, now $\sim$10\,000), and the photometric binary
investigation by Milone et al.\ (2012a), will allow us to survey
different types of binaries and to constrain their spatial and period
distributions, over the whole central region of M\,4.
Because the heavy binary properties in the theoretical models are
\textit{almost entirely unconstrained} by existing observations, we
expect that the comparison between theory and observation will reveal
differences with the assumed properties of the binary population.
Further modeling will be conducted to reconcile these (see
Section~\ref{dyn_mod}).

\begin{figure*}
\includegraphics[width=170mm,height=85mm]{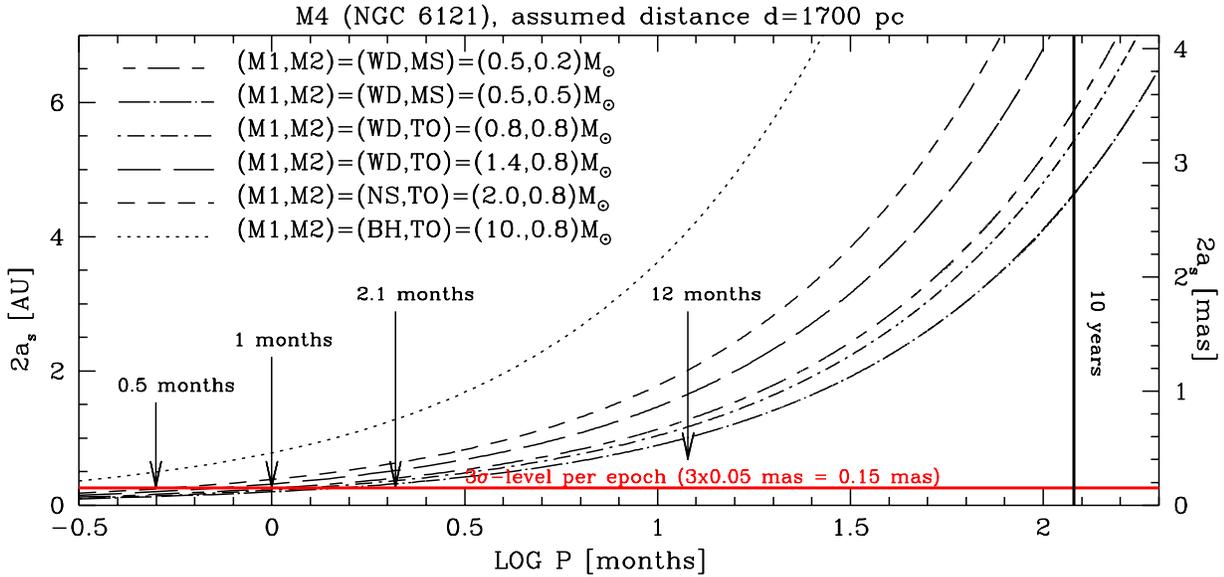}
\caption{
The relation between twice the semi-major axis 2$a_{\rm s}$ for the
barycentric orbit of the luminous companion (because that is the
maximum size of the total range of motion for a circular orbit that we
can measure) and period (P) for binary stars of different masses. Our
observing strategy should detect binaries with a minimum period of
$\sim$1 month.
The horizontal red line sets our
limit, due to the resolution of the telescope and the precision of our
method (3$\sigma$ $\simeq$0.15 mas, i.e., $a_{\rm s}$ $\sim$0.1 AU for a
distance of 1700 pc), while the vertical line to the right sets the
longest detectable period ($\sim$10 years) making use of the existing 
\hst\/ archival material. 
}
\label{mm}
\end{figure*}

\section{The target}
\label{tgt}
M\,4 is our target of choice for several
reasons:\ 
(1) 
as explained above, we expect to find a high concentration of binaries;
(2) the cluster core is large and open, and we expect most of the
massive binaries to have settled there, due to mass segregation;
(3) M\,4 is nearby, which allows study of binaries below the turn-off 
(i.e., more stars); 
(4) the closest GC is the place where the angular wobbles will be
   the largest.
The expectation that a relatively large fraction of stars in the core
of M\,4 are in binaries is confirmed by the work by Milone et
al.\ (2012a), where, using a color-magnitude diagram (CMD) for the
inner 50\arcsec, it was found that about 15\% of the proper-motion
members are above and to the red of the main sequence (MS).  These
must be MS+MS binaries.
There should also be a population of binaries with a MS star and a
massive, evolved companion [white dwarf (WD), neutron star (NS),
  possibly stellar-mass black hole (BH)] that cannot be detected
photometrically.

Binary systems with massive remnant components must exist, simply from
normal stellar evolution.  Furthermore, simulations show that MS+MS 
binaries should tend to exchange low-mass MS companions for more
massive ones, including massive WD and NS.
A few heavy binaries in M\,4 have indeed been revealed through their
X-ray emission (Bassa et al.\ 2004), but the vast majority remain
undetected.
[Note also, that M\,4 certainly hosts a pulsar, in a triple system with a
WD and a planetary-mass companion, at $\sim$1 core radius $(r_{\rm c})$
(Sigurdsson et al.\ 2003)].
%
%
Our study is thus important for constraining not only the fraction of
binaries and their properties, but also the amount of unseen mass in
GCs (including the raw ingredients for the production of binaries with
compact objects).


\section{Expected properties of the binaries}
\label{expected}
Binaries evolve both dynamically and through their own internal
evolution. We expect that the softest primordial binaries (semi-major
axis, $a$ $>$ 2 AU) have almost all been destroyed.  According to our
current best model (Heggie \& Giersz, 2008) the distribution of the
semi-major axes for the present-day binaries is expected to have a
median value of $a\sim$0.3 AU, with the bulk of the binaries having
$a$ between 0.05 and 1.2 AU.
If we want to detect binaries we need to be sensitive to the wobbles
of such systems.

Our models also give the period, luminosity and spatial distribution
of the binaries, and are consistent with the available information on
photometric binaries and radial velocity binaries (in the outskirts).
The models also include {\sl exchange interactions} (Hills 1976), the
main route by which degenerate binaries form.  However the cross
sections which were used for these processes are approximate, and so
we \textit{also} give the following more model-independent argument 
to estimate the number of heavy binaries.

To obtain an estimate of the expected number of detectable binaries,
we argue as follows ({\em bearing in mind that the aim of this project
  is to provide the basic data on which such estimates depend!}).  In
the core of the Monte Carlo model there are $\sim$1800 white dwarfs
(41\% of all objects in the core, in agreement with the observations
from GO-10146, see Bedin et al.\ 2009).  Of these, $\sim$1200 have
masses above the mass of stars at the MS turn-off
($\sim$0.8$M_\odot$).  The model also predicts that at the present day
there should be up to 1200 neutron stars within the core (depending on
the retention fraction).
It has been estimated (Ivanova et al.\ 2008) that of order 50\% of NS
stars in the core would form binaries by exchange, in a GC with 100\%
primordial binaries.  Therefore we expect that, in the core of M\,4,
where the binary fraction is about 10\%, the number of binaries with
WD or NS companions could be as much as $(1200+1200) \times 0.50
\times 0.10 \simeq 120$, and no less than 60 with WD companions alone.
According to our model, about 50\% of the binaries in the core lie
less than $\sim$4 magnitudes below the turn-off, and so the expected
number of such binaries with WD or NS companions should be in the
range $\sim$30--60.  Reassuringly, the Monte Carlo models predict even
larger numbers: $\sim$100.


\section{Detecting the wobble} 
\label{detection}
When \textit{Wide-Field Camera~3} (WFC3) was installed on
\textit{HST\,} during Servicing Mission~4 (SM\,4), we were very
excited about the astrometric potential of the Ultraviolet and Visual
(UVIS) channel. With a field close to that of the Wide-Field Channel
(WFC) of the \textit{Advanced Camera for Surveys} (ACS), and smaller
pixels (0.039775 arcsec) with less Charge Transfer Efficiency (CTE)
losses (to start with), WFC3/UVIS would clearly be the instrument of
choice for imaging astrometry.  When the Servicing Mission Orbital
Verification (SMOV) images were released, we immediately set out to
apply the expertise we have gained working with the ACS detectors
(Anderson \& King 2004a and 2006, hereafter AK) and solved for the
distortion and demonstrated its astrometric prowess (Bellini \& Bedin
2009).  Since SMOV, many more astrometric calibration images have been
taken through many more filters, and we have recently published a
distortion solution for the ten most commonly used filters that is
accurate to better than 0.01 pixel (Bellini, Anderson, \& Bedin 2011).
The solution accounts for the camera optics, filter-specific
perturbations, and a $\pm$0.03-pixel manufacturing defect.  Once the
distortion has been removed, we are able to recover the full
astrometric capabilities of UVIS.
Bellini et al.\ (2011) demonstrates the measurement quality for a
single UVIS exposure as a function of instrumental magnitude.  For
stars with good S/N, we achieve a single-measurement accuracy of
0.0085 pixel, or about 0.35 milli-arcsec (mas), along each coordinate
(as confirmed also by our preliminary reduction of the first epoch;
see Sect.~\ref{1ste}).

The core radius of M\,4 is $r_c$$\sim$50$''$, and therefore the
WFC3/UVIS field will cover the entire core region to detect many kinds
of binary systems, with various masses and periods, with typical
examples summarized in Fig.~\ref{mm}.  The range of separations that
we will be able to measure corresponds to a range of periods,
depending on the component masses.
The horizontal red line at $\sim$0.15 mas marks where
our 3-$\sigma$ astrometric detection limit will be.  

The above-mentioned measurement precision in a single image (of
$\sim$0.35 mas) for each star can be improved by taking many
independent exposures at a variety of dithered positions within each
epoch, and by adopting a local-transformation approach (Bedin et
al.\ 2003; 2006; Anderson et al.\ 2006; Milone et al.\ 2006; Anderson
\& van der Marel 2010).  The astrometric errors at each epoch scale as
$\sim 1/\sqrt N$, where $N$ is the number of images on which the star
is measured.  The dithering and the differential-astrometric
approaches allow us to mitigate any small errors in the solution for
the distortion, which is surely present in images, following the
procedures we have already applied to the other cameras (WFPC2, HRC
and WFC:\ AK03a, AK04b, AK06).  With $\sim$50 observations of each star
per epoch, our ultimate 1-$\sigma$ precision can reach $\sim$0.05 mas
(i.e.\ $\sim$1.2 milli-pixel).

\begin{figure*}
\includegraphics[width=60mm,height=60mm]{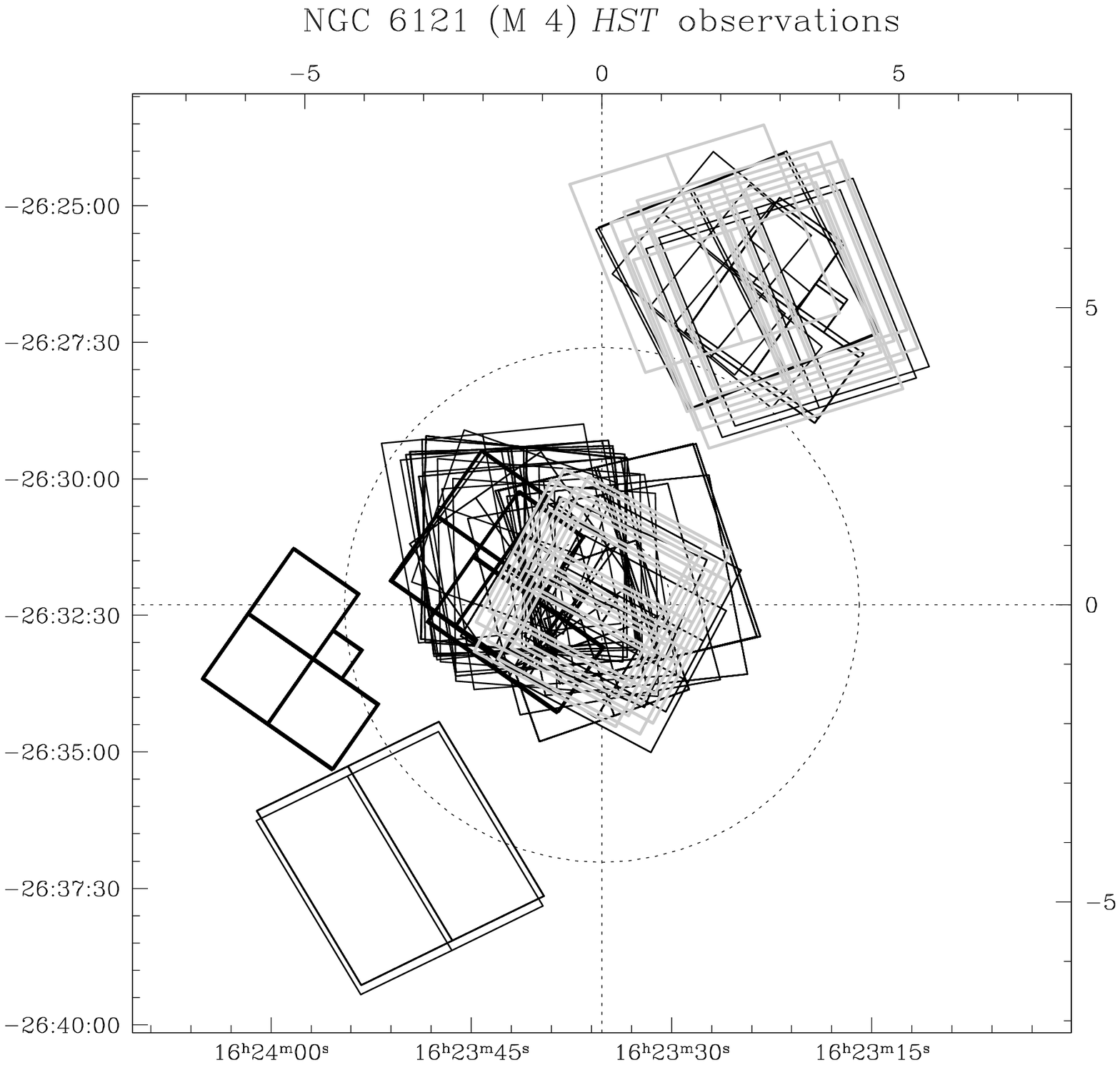}
\includegraphics[width=9mm, height=60mm]{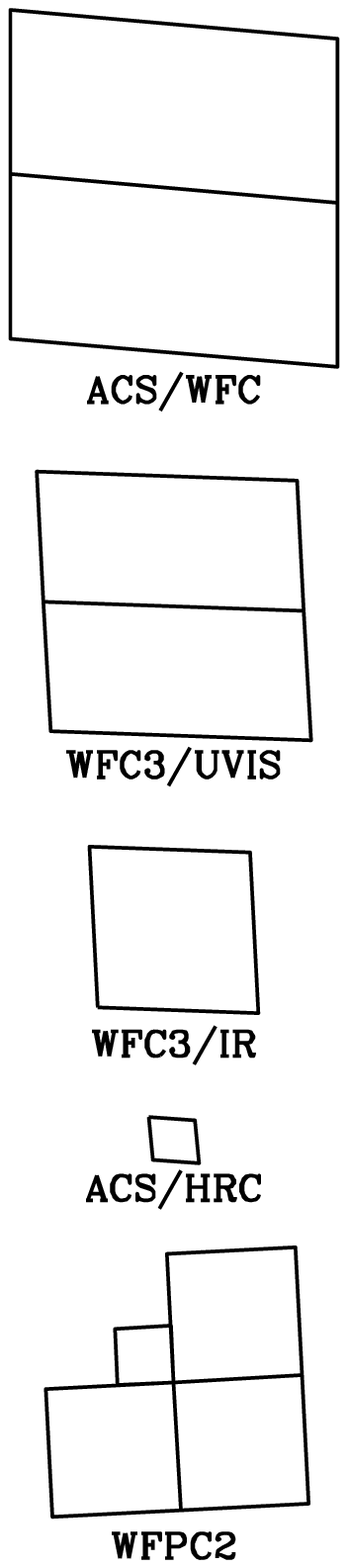}
\includegraphics[width=40mm,height=60mm]{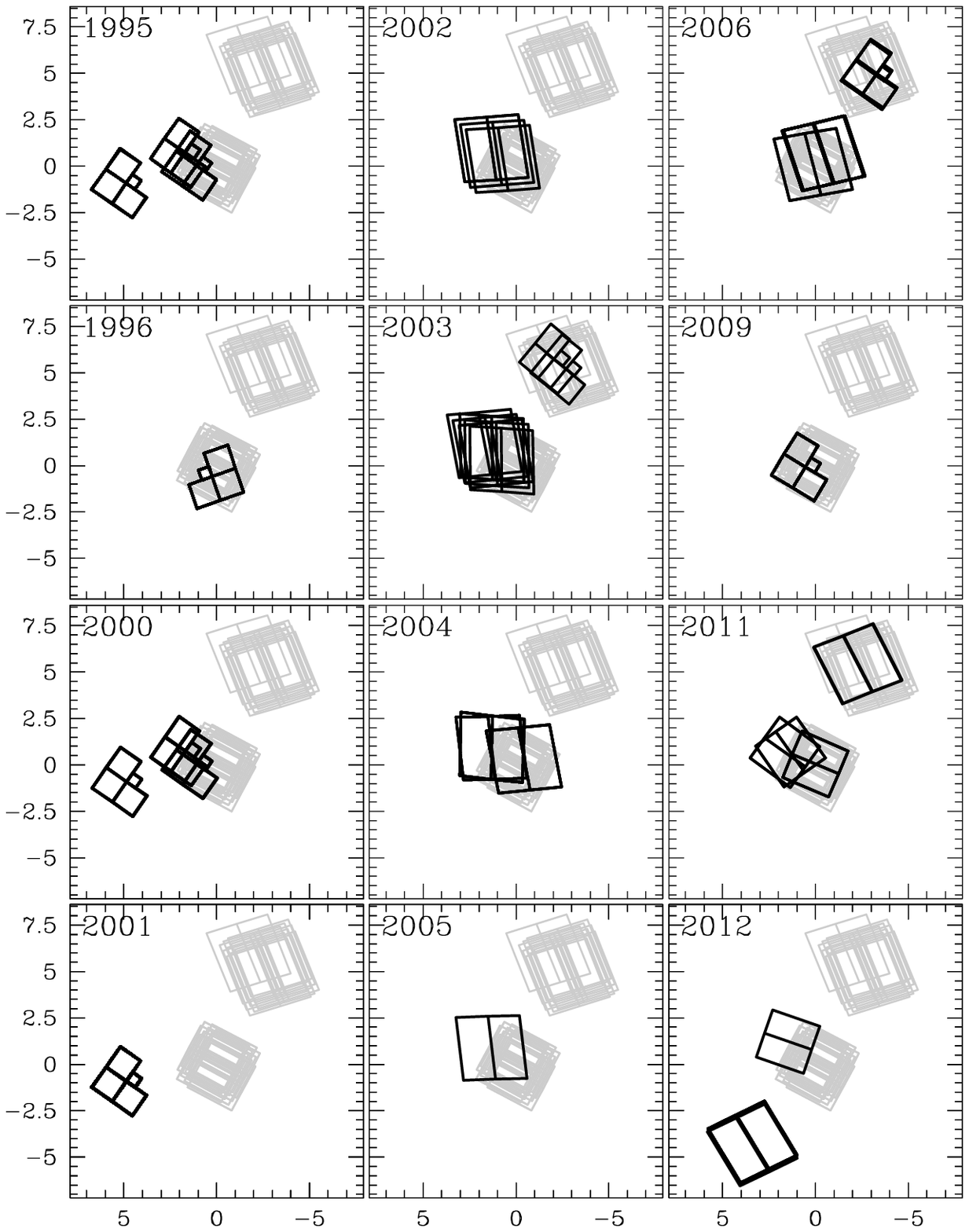}
\includegraphics[width=60mm,height=60mm]{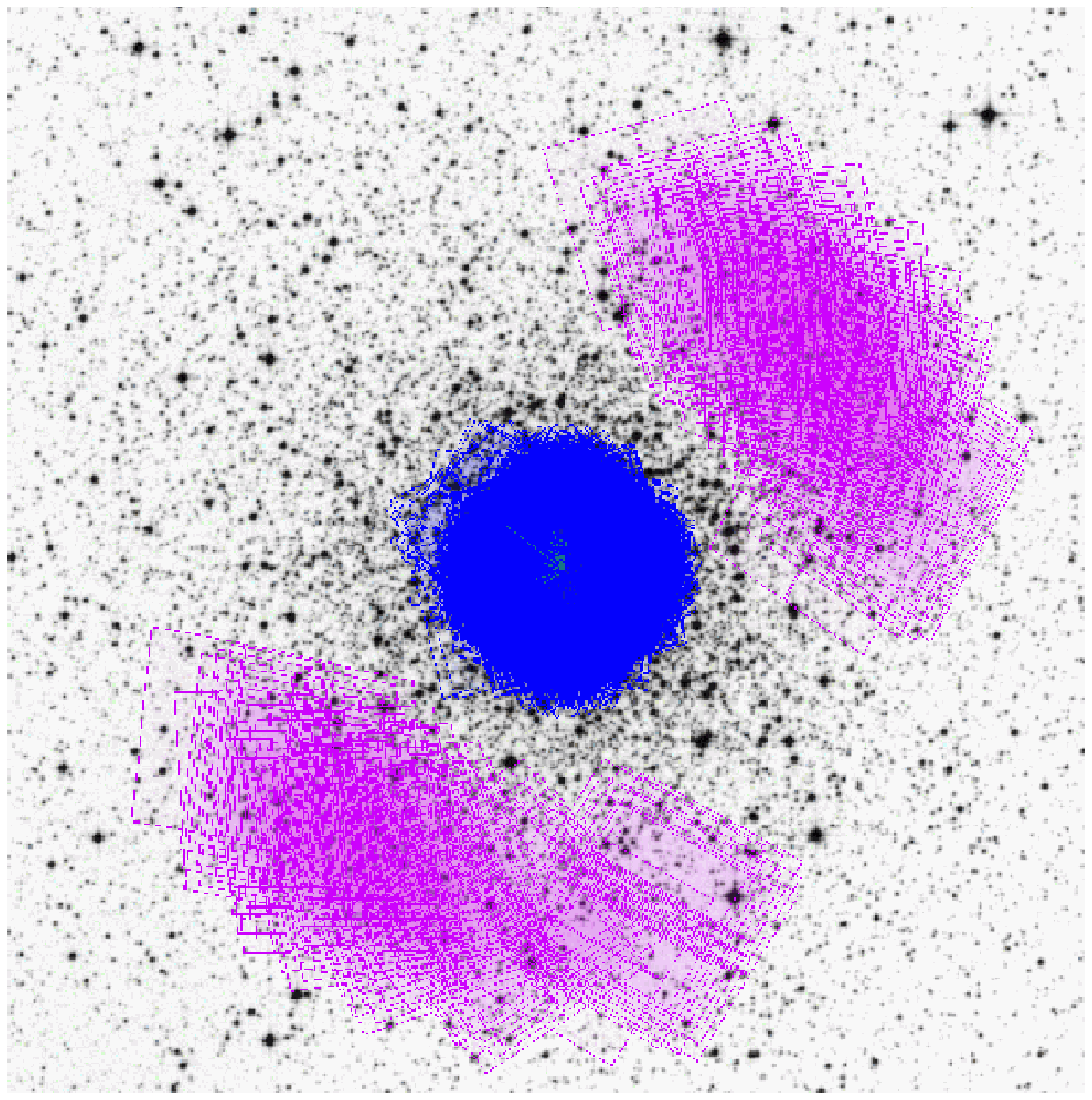}
\caption{
\textit{(From Left to Right\/):}
The first panel shows all the existing \hst\/ observations available
in the M\,4-field (as October 2012).  The $\alpha$ and $\delta$
(J2000.0) are indicated on bottom and left axes, while top and right
axes give the angular distance in arcmin from the assumed center of
M\,4. Dotted lines mark the center of the cluster, and the 5 arcmin
angular distance. Data from the first visit of GO-12911 are
indicated
in grey.  On the right of this panel, for reference, are shown on the
same scale the footprints of the different detectors.
The second panel shows in a 3$\times$4 array the archival material 
subdivided per year (labeled on the top-left of each box). 
The last panel shows, overlaid to a DSS image of M\,4, the precise locations of
all the planned individual images for GO-12911 (WFC3/UVIS in blue,
ACS/WFC parallels in magenta, from the \hst\/'s \textit{Astronomer's
  Proposal Tool}\,).
}
\label{obs}
\end{figure*}

To put this precision into context, let us consider a massive binary
made of a NS (of 2 $M_\odot$) and a MS star (of 0.6 $M_\odot$) with
period of $\sim$15 days. The orbit of the luminous component, the MS
star (the {\it secondary)}, around the center of mass of the pair,
would have a semi-major axis ($a_{\rm s}$) of $\sim$0.13 AU.
At the distance of M\,4 ($\sim$1.7\,kpc, Peterson et al.\ 1995) this
implies a semiamplitude of the wobble of $\sim$0.075 mas. We note, in
fact, that over its orbit such a star will show a total range of up to
twice the barycentric semi-major axis, bringing its total excursion to
2$\times$0.075 mas = 0.15 mas, our 3-$\sigma$ detection limit.
All the binaries above the red line in Fig.~\ref{mm} can be discovered by our
survey.  As the period increases, the wobble amplitude increases,
making the binary easier to detect.  
Data of our large program GO-12911 will be sensitive to binaries with
periods up to $\sim$1-year.  Although we expect only about 25\% of
binaries to have periods larger than $\sim$1 year, by including
observations from the archive taken 4, 6, 10, and 15 years before
Cycle-20, we could explore much longer periods (see
Fig.~\ref{obs}). The lower precision of some of the archival epochs
might be compensated for by the larger wobble of the longer periods.

Measuring the wobble of a star implies also solving for the motion of
the binary system with respect to a members-rest-frame.  Solving for
the members-rest-frame and for the individual motions of the
binary system luminous components is an iterative process.

From archival images, and as later confirmed by our first epoch, we
know that there are $\sim$2000 MS stars for which we can measure
high-precision astrometric positions.
Figure~\ref{mm} shows that the detectable binaries will have
periods longer than 1 month, and our model shows that over 50\% of the
bright binaries in the core should have periods at least as large.
Combining this with the expected number of degenerate binaries in the
core, we therefore expect the survey to allow us to discover
$\sim$15-30 such binaries ($\sim$30-60$\times$0.50).
Our search will yield estimates of both the period and semi-major
axis, and we can therefore estimate the total mass in the system.
Knowing the luminosity of the MS star we can estimate its mass, 
then get the mass ratio $q$, and so the mass of the dark companion.
The numbers and properties of these binaries are the goal of this
project.

Note that GAIA will not be able to accomplish the goals of this
project because of crowding, neither will JWST or ground-based AO/MCAO
systems, because they operate in the near-IR where the halos of the
bright red giants will interfere with the measurement of our target
upper main-sequence stars.

\section{Observing strategy}
\label{obsstg}
In planning the observing strategy, we had three main priorities:
\begin{itemize}
\item we wanted to maximize the number of stars in the core 
          of M\,4 with suitable S/N; 
\item we wanted to obtain images with the sharpest PSFs and 
          the best possible sampling and stability;
\item we wanted to maximize the number of independent exposures. 
\end{itemize}

\subsection{Choice of the camera} 
To maximize the number of stars we need to use a detector that covers
the entire core of M\,4 ($r_{\rm c}$$\sim$$50''$).  This restricts our
choice to either ACS/WFC or WFC3/UVIS in full-frame mode, as
subarray-mode (which would have a better duty-cycle) is not a suitable
option.
We also need to suppress the light from red giants in the core of the
cluster so that their wide PSF halos will not contaminate the
measurement of the other stars.  This can be done by using a blue
filter (see next section).
The choice was to use WFC3/UVIS, which has a better pixel sampling
than ACS/WFC, is optimized for blue light, and has the additional
advantages explained below.

\subsection{Choice of filter}
WFC3/UVIS (like ACS/WFC) has an optimal exposure-time of $\sim$350 s
which minimizes the overhead and maximizes the duty-cycle.\footnote{
Above this threshold, the successive exposures can be collected during
the downloading of the previous ones, below this threshold the
telescope will only perform the downloading.
}
At the same time, to reduce our systematic and random errors (see
discussions in Sect.~\ref{dith} and \ref{orient}) we need the largest
possible number of independent exposures.  For these reasons we would
take as many images at the minimum-optimal length as possible
($\geq$350 s).

Just above the sub-giant branch (SGB) is where the LF takes a sharp
upturn, and so using a filter that saturates just above this limit
yields the largest number of stars within 4 magnitudes of saturation,
where we have optimal S/N.
Among the broad-band filters, with these kinds of exposure-times,
F390W is the only one that saturates stars just brighter than the SGB,
and for this reason, this filter was initially selected at the
phase\,I stage of the proposal.

However, at the phase\,II stage ---thanks also to archival and
proprietary material that just became available--- we realized that
the PSFs in filter F390W (and in all filters bluer than $\sim$400nm)
were not good.
Not only did the positions turn out to be significantly
color-dependent (as already demonstrated by Bellini et al.\ 2011), but
PSFs below $\sim$400nm are also considerably wider and less stable
than in redder filters.

In the end, we selected the medium-band filter F467M, since archival
images showed that it saturated at the same level as F390W, but the
PSF is 10\% sharper, which translates directly into 10\% better
positions.  An additional benefit of the medium-band filter is that it
significantly suppresses any color-dependent effects, which are
clearly seen in WFC3/UVIS's bluer filters.

\subsection{Number of exposures per epoch}
\label{nee}
In each orbit, we were able to fit 5 exposures of $\sim$390 s depth
and one exposure of 20\,s through F775W at the beginning of each
orbit.  There was no way to trade the F775W exposures for more F467W
exposures, and the F775W exposures give us a handle on color issues.
This strategy yields the highest number of well-exposed images of
stars around the MS turn-off.  Our images from the archive show that
stars in this field are sufficiently isolated to be measured
accurately as single stars.  At the same time, there will be plenty of
nearby high S/N neighbors so that we can measure each star's position
relative to a local network of its neighbors, thereby minimizing
distortion errors.  On average each star will be $\sim$30 pixels from
any neighbor, but there will be $\sim$100 high S/N neighbors within
$\sim$500 pixels (i.e.\ $\sim$20$''$).
%
In Sect.~\ref{1ste}, we show that the first epoch achieves an astromeric
precision of $\sim$0.008 pixel ($\sim$0.35 mas) per coordinate.  With
50 exposures in each ten-orbit epoch, this gives us a 1-$\sigma$ precision 
of 0.05 mas per epoch, which is our target accuracy for detecting the 
wobbling binaries.

\begin{figure}
\includegraphics[width=80mm,height=80mm]{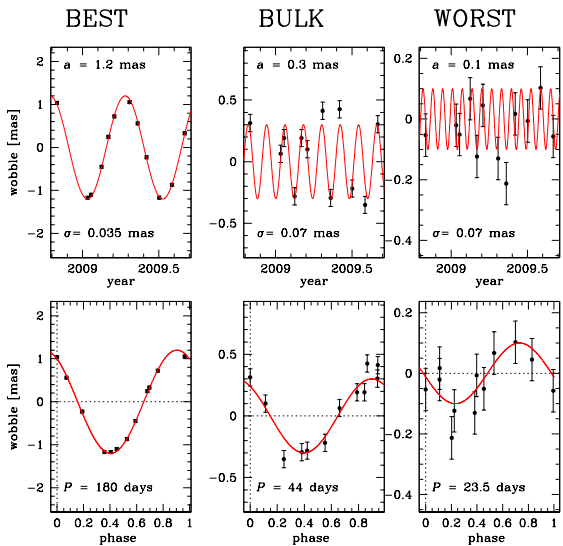}
\caption{Examples of the expected wobbles.  In three representative
  cases of detectable wobbles (red lines), we show how single-epoch
  simulated data-points (black dots) would look, assuming the
  barycentric semi-major axes $a$ (indicated on top of top panels) and the
  uncertainties $\sigma$ (on bottom of top panels). Bottom panels show
  the phased signals and label the periods.  }
\label{wobbles}
\end{figure}

\subsection{Number of epochs} 
The problem of computing orbital elements of a binary from a set of
observed positions is formally analogous to the case of orbits in the
solar system, and yet in practice there is little resemblance between
the methods employed.  In particular measurement errors are much
larger (compared to the quantities to be determined) in the
binary-star case, so that an orbit needs to be based on a much larger
number of observations.

There are several methods for the determination of visual binary
orbits (see Heintz 1978 for a review).  In our case we are dealing
with pure astrometric orbits, computed \textit{(1)\,} from 2-D
position measurements, and \textit{(2)\,} with no a priori information
about the orbit (i.e.\ no RV orbit).  Under these assumptions, and
with our expected S/N ratio (errors of 0.05 mas on a separation of
0.20-0.50 mas, for the bulk of the binaries), simulations (such as
those in Fig.~\ref{wobbles}) show that an orbit derived from
\textit{twelve} 2-D positions well distributed in phase over at least
two binary periods should provide reliable detections.  The minimum
stipulation of two periods is important for confirming that our model
for the tightest detected binaries ($P$$<$6 months) is predictive, and
for a realistic estimation of errors.  In the case of {\em
  non-circular\/} orbits it is likely that twelve measurements would
determine $a$ (or at least the maximum separation) considerably better
than other elements such as the eccentricity.
For all the above reasons, we divided our 120 orbits into twelve
10-orbit epochs separated by a minimum of $\sim$15 days to a maximum
of $\sim$3 months.  This sampling should allow us to probe binaries
with periods between $\sim$15 days to 6 months. 

\subsection{Dither pattern}
\label{dith}
For each epoch (of 50 images) we planned to use a grid of
$\sim$7$\times$7 pointings with a $\sim$200-pixel step. 
All images will be dithered as described in AK00. 
The dither strategy is fundamental for the 
success of this program.  It will be crucial:
(1) to solve for the residual geometrical distortion within each 
    epoch; 
(2) to solve for the effective ePSF of the undersampled detector within
    each epoch; 
(3) to have a handle on any unaccounted sources of systematic error
    (CTE inefficiencies, detector and optical artifacts, chromatic
    effects, etc...).
Our choice of strategy gives the maximum dither (thus optimizing the
randomization of any unremoved distortion error) while also satisfying
the constraint of covering the entire core of M\,4 in every exposure.

\subsection{Orientations}
\label{orient}
In addition to having a large number of pointings, each epoch will
have a different orientation.  This will allow us to construct the
most accurate possible average distortion correction.  While
distortion errors will be mitigated by our local transformations, it
is always advantageous to start with the most accurate possible global
solution.  Again, this will allow us also to remove any unanticipated
(and anticipated) systematic effects, while giving us the entire inner
$\sim$1{\hbox{$.\!\!^{\prime}$}}8 at every visit (essentially the
entire core of the cluster, see right panel of Fig.~\ref{obs}).

\subsection{CTE mitigation} 
\label{ctem}
In filter F467M we have a relatively low background.  Although our
focus is on the bright stars, in this situation CTE might potentially
introduce astrometric systematic errors at the fainter magnitudes.
WFC3/UVIS is still young, but it has suffered more CTE damage than ACS/WFC
had at its age.  In Anderson \& Bedin (2010), we designed the
pixel-based correction for CTE in ACS/WFC images. The Space Telescope
Science Institute is exploring the best way also to deal with CTE in
UVIS data.  With our observing strategy, we should have an excellent
empirical handle on the impact of CTE and are confident that it will
not appreciably impact our results.
To further mitigate the impact of CTE inefficiencies, we have also
imposed a post-flashing to all our WFC3/UVIS (see
Sect.~\ref{Swfc3uvis}).

\section{Previous large surveys of GCs with  \hst}
\label{cmp}
There have been four previous \hst\/ Large Programs on globular clusters.
It is interesting to compare those archival data-sets with our
GO-12911 to see what this new program will add to the \hst\/ legacy.

We describe the previous programs in chronological order.  
GO-8267 (PI: Gilliland) used WFPC2 to observe the core of 47\,Tuc.
The observations were collected with a differential-photometry
strategy in mind, and therefore no intentional dither is
present. These data were collected in less than 10 consecutive days to
be sensitive to exo-planets of Jupiter size with periods shorter than
that (Gilliland et al.\ 2000).
GO-8679 (PI: Richer) also occurred in the pre-ACS era, using WFPC2 to
observe an outer field of M\,4, attempting to reach the end of the
white dwarf cooling sequence.  Data were mildly dithered, and
observations spanned about three months (Richer et al.\ 2002).
GO-10424 (PI: Richer) and GO-11677 (PI: Richer) are similar projects
to the previous one.  They are both aimed at the study of the faintest
stars of the clusters NGC\,6397 (Richer et al.\ 2006) and 47\,Tuc
(Kalirai et al.\ 2012), respectively.  They both studied a field in
the outskirt of the clusters, and both made use of ACS/WFC.  In the
case of GO-10424, there are three deep images per orbit, and they were
all collected in about a month. In the case of GO-11677, there are
just two deep $\sim$1400\,s exposures per orbit, collected in eight
months.
In summary, among these archival large programs, only GO-8267 observed
the core of a globular cluster and focused on the same type of stars
as our GO-12911, i.e., the four brightest magnitudes just below the
MSTO.  These facts make GO-8267 the most similar data-set to GO-12911
in the \hst\/ archive, and yet there are significant differences
between the two programs.
The most important difference is that GO-12911 images are collected in
twelve epochs spanning one year instead of being consecutive. Second,
GO-12911 images are planned to be dithered within each given epoch,
which naturally worsens the differential photometric precision, but
does enhance photometric accuracies, and ---most importantly--- it is
a necessity for high accuracy astrometry.
Exposure cadence ($\sim$390 s) is significantly lower in GO-12911 than
in GO-8267 (160 s), but is significantly higher than in the three
other Large Programs on GCs (GOs: 8679, 10424, and 11677).
Interestingly, all these programs collected color information during
each orbit.

GO-12911 is the first \hst\/ large program to observe the core of a
globular cluster using a new generation instrument such as WFC3/UVIS
and the first of the four to collect observations with large dithers.


\section{First epoch}
\label{1ste}

The first of the twelve epochs for GO-12911 was collected on October
9th and 10th 2012.
In this section we provide the astronomical community with a report
about this data-set, which should be considered as a template for what
is expected from each of the remaining epochs.  To this end, we
illustrate the photometric and astrometric quality achieved in our
pre\-li\-mi\-na\-ry reductions.

\subsection{Observations} 

The first epoch (hereafter, \textit{epoch\,01}\/), as well as the next
eleven planned, consists of ten single-orbit \hst\/ visits, during
which WFC3/UVIS is the primary instrument and ACS/WFC is used in
parallel.  All images within a given epoch are taken with the same
\hst\/ roll-angle; in the case of epoch\,01, this was set to be
$\textsf{ORIENT}=107^\circ$ (i.e., PA\_V3$\simeq$286$^\circ$).

During each orbit, \hst\/ takes 1 short exposure of 20\,s in
WFC3/UVIS/F775W, 5 astrometric images of $\sim$390\,s in
WFC3/UVIS/F467M, and a single coordinated parallel ACS/WFC exposure of
180\,s, which is alternately taken through F814W and F606W (only one
parallel exposure could be taken each orbit on account of buffer-dump
limitations).  One of the short WFC3/UVIS exposures, at the center of
the dither pattern, was chosen to be collected through F467M rather
than in F775W, and at the same location as the subsequent F467M long
exposure.
This will enable us to register our astrometry and photometry for
saturated stars to positions and fluxes of the unsaturated objects.
As described in Sect.~\ref{dith}, our astrometric WFC3/UVIS/F467M
exposures are organized in a 7$\times$7 dither pattern, and, since in
10 orbits we can fit 50 images, we have one extra astrometric long
exposure which breaks this symmetry.  We chose to place this extra
exposure on the background extragalactic point source described in
Bedin et al.\ (2003) using either F467M or F775W.  This way we will
have a cross-check on the absolute zero-motion and an additional
check on the geometric distortion in the outskirts of our studied core
field.

\begin{figure}
\includegraphics[width=80mm,height=80mm]{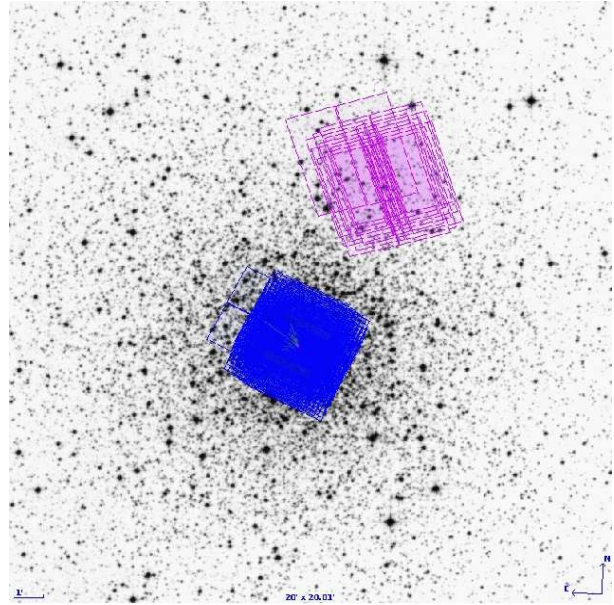}
\caption{
Layout of the observations collected during epoch\,01 of M\,4.  The
simultaneous coordinated parallel ACS/WFC images are shown in magenta,
while the primary WFC3/UVIS images are shown in blue (from \hst\/'s
\textit{Astronomer's Proposal Tool}\,).
}
\label{fov01}
\end{figure}

Figure~\ref{fov01} shows the location of the footprints of the individual 
images collected during epoch\,01 superposed on a Digital Sky Survey image, 
and Table~\ref{tab01} summarizes the data-set.

To collect images in conditions as identical as possible, we plan to
collect all the images, in a given epoch, within no more than three
consecutive days (this was successfully done for the first 6 epochs).
Tentative scheduling and roll-angles for the next visits are given in
Table~\ref{tab02}.

\begin{table}
\caption{Data-set summary for 
  epoch\,01 of program
  GO-12911. Visits: \textsf{0L,0M,0N,0O,0P,0Q,0R,0S,0T}, and \textsf{0U} were
  collected between October 9$^{\rm th}$ 2012 at UT 00:49:56 and October
  10$^{\rm th}$ 2012 at UT 22:01:47. }
\label{tab01}
\begin{tabular}{ccc}
\hline
Camera/Channel & Filter & \#$\times$Exposure-time\\ 
\hline 
\multicolumn{3}{c}{ \textit{ primary observations} } \\
WFC3/UVIS         & F467M  &  1 $\times$  20 s \\  
WFC3/UVIS         & F467M  & 45 $\times$ 392 s \\  
WFC3/UVIS         & F467M  &  5 $\times$ 396 s \\  
WFC3/UVIS         & F775W  &  9 $\times$  20 s \\  
\hline
\multicolumn{3}{c}{ \textit{coordinated parallel observations} } \\
ACS/WFC           & F606W  &  5 $\times$ 180 s \\
ACS/WFC           & F814W  &  5 $\times$ 180 s \\
\hline
\end{tabular}
\end{table}


\begin{table}
\caption{Plans for scheduling and \textsf{PA\_V3} of all epochs. }
\label{tab02}
\begin{tabular}{ccc}
\hline
Epoch ID & Date & \textsf{PA\_V3}[$^\circ$] \\ 
\hline 
{\bf 01}$^\dagger$ & {\bf 09-10 Oct.\ 2012} & {\bf 286} \\
{\bf 02}$^\dagger$ & {\bf 18-19 Jan.\ 2013} & {\bf  92} \\
{\bf 03}$^\dagger$ & {\bf 11-12 Feb.\ 2013} & {\bf  99} \\
{\bf 04}$^\dagger$ & {\bf 07-08 Mar.\ 2013} & {\bf  83} \\
{\bf 05}$^\dagger$ & {\bf 01-02 Apr.\ 2013} & {\bf 110} \\
{\bf 06}$^\dagger$ & {\bf 24-25 Apr.\ 2013} & {\bf 124} \\
     07           &     18-20 May.\ 2013  &       150  \\
     08           &     11-13 Jun.\ 2013  &       240  \\
     09           &     05-07 Jul.\ 2013  &       253  \\ 
     10           &     29-31 Jul.\ 2013  &       260  \\
     11           &     22-24 Aug.\ 2013  &       267  \\
     12           &     15-17 Sep.\ 2013  &       277  \\
\hline                  
\multicolumn{3}{l}{
$^\dagger$ Already collected. 
}
\end{tabular}
\end{table}

\subsection{Data reduction and calibration} 

While reduction procedures for WFC3 are still being improved, the
procedures for point-source astrometry and photometry in ACS/WFC
images have already reached a point where any future improvement is
likely to be marginal.
Therefore, we will first discuss the less-preliminary reductions
obtained for the coordinated parallel ACS/WFC observations, and we
conclude the presentation of the new data with the more-preliminary
reduction available (at this stage) for the WFC3/UVIS data.

\subsubsection{ACS/WFC} 

All CCD detectors in the harsh radiation environment of space suffer
degradation due to the impact of energetic particles, which displace
silicon atoms and create defects.  These defects can temporarily trap
electrons during readout, resulting in Charge Transfer Efficiency
losses and in trailing of the sources (because the electrons are
released at a later time). These effects have a major impact on
astrometric projects (Anderson \& Bedin 2010) and also non-negligible
effects in photometry.  In this work, every single ACS/WFC image
employed was treated with the pixel-based correction for imperfect
CTEs developed by Anderson \& Bedin (2010).  The algorithm has been
now improved\footnote{
    The pixel-based CTE correction scheme based on the work of Anderson 
    \& Bedin (2010) has been modified to include the time- and 
    temperature-dependences of CTE losses (Ubeda \& Anderson 2012). 
    An improved correction at low signal and background levels has also 
    been incorporated, as well as a correction for column-to-column 
    variations. 
}
and is now a part of the standard CALACS pipeline as \_{\sf flc}
exposures.  This correction has been shown to be effective in
restoring fluxes, positions, and the shape of sources, and reducing
systematic effects of imperfect CTE to less than $\sim$20\% except in
the case of extremely low backgrounds (Anderson \& Bedin 2010, Ubeda
\& Anderson 2012).

Fluxes and positions of sources within ACS/WFC images were extracted
with the software tools described in detail by Anderson et
al.\ (2008).  It consists of a package that analyzes all the exposures
simultaneously in order to generate a single list of stars.  This
routine was designed to work well in both uncrowded and moderately
crowded fields, and it is able to detect almost every star that can be
detected by eye. It takes advantage of the many independent dithered
pointings of the images and of the knowledge of the PSF to avoid
including artifacts in the list.  We used spatially-variable library
effective PSFs documented in AK06, but the parallel fields are too
sparse to allow us to improve the PSF with the software's
``perturbation'' option.

The photometry was calibrated into the WFC/ACS Vega-mag system
following the procedures given in Bedin et al.\ (2005), and using
the most updated encircled energy and zero points 
available at the STScI's ACS web-site.\footnote{
http://www.stsci.edu/hst/acs/analysis/zeropoints
}
We will use for these calibrated magnitudes the symbols $m_{\rm F606W}$
and $m_{\rm F814W}$.

Finally, stars that saturate (brighter than $m_{\rm {F606W}}$ $\sim$
18.5 and $m_{\rm {F814W}}$ $\sim$ 17.5) are treated as described in
Sect.\ 8.1 in Anderson et al.\ (2008).  Collecting photo-electrons
along the bleeding columns allows us to measure magnitudes of
saturated stars up to $\sim$3.5 mag above saturation (i.e.\ up to
$m_{\rm {F606W}}$ $\sim$15, and $m_{\rm {F814W}}$ $\sim$ 14), with
errors of only a few percent (Gilliland 2004). This treatment allows
us to recover the MS TO stars also in the ACS/WFC parallel images of
the outskirts of M\,4.

With the transformations from the coordinates of each image into a
common frame, it becomes possible to create a stacked image of the
field within each epoch.
The stack provides a representation of the astronomical scene that
enables us to independently check the region around each source at
each epoch.
The stacked images are 15\,000 $\times$ 15\,000 super-sampled pixels
(by a factor $\sim$2 relative to the image pixels, i.e.\ $\sim$25
mas/pixel), corresponding to $\sim$6$\times$6 arcmin$^2$.

\begin{figure*}
\includegraphics[width=55mm,height=55mm]{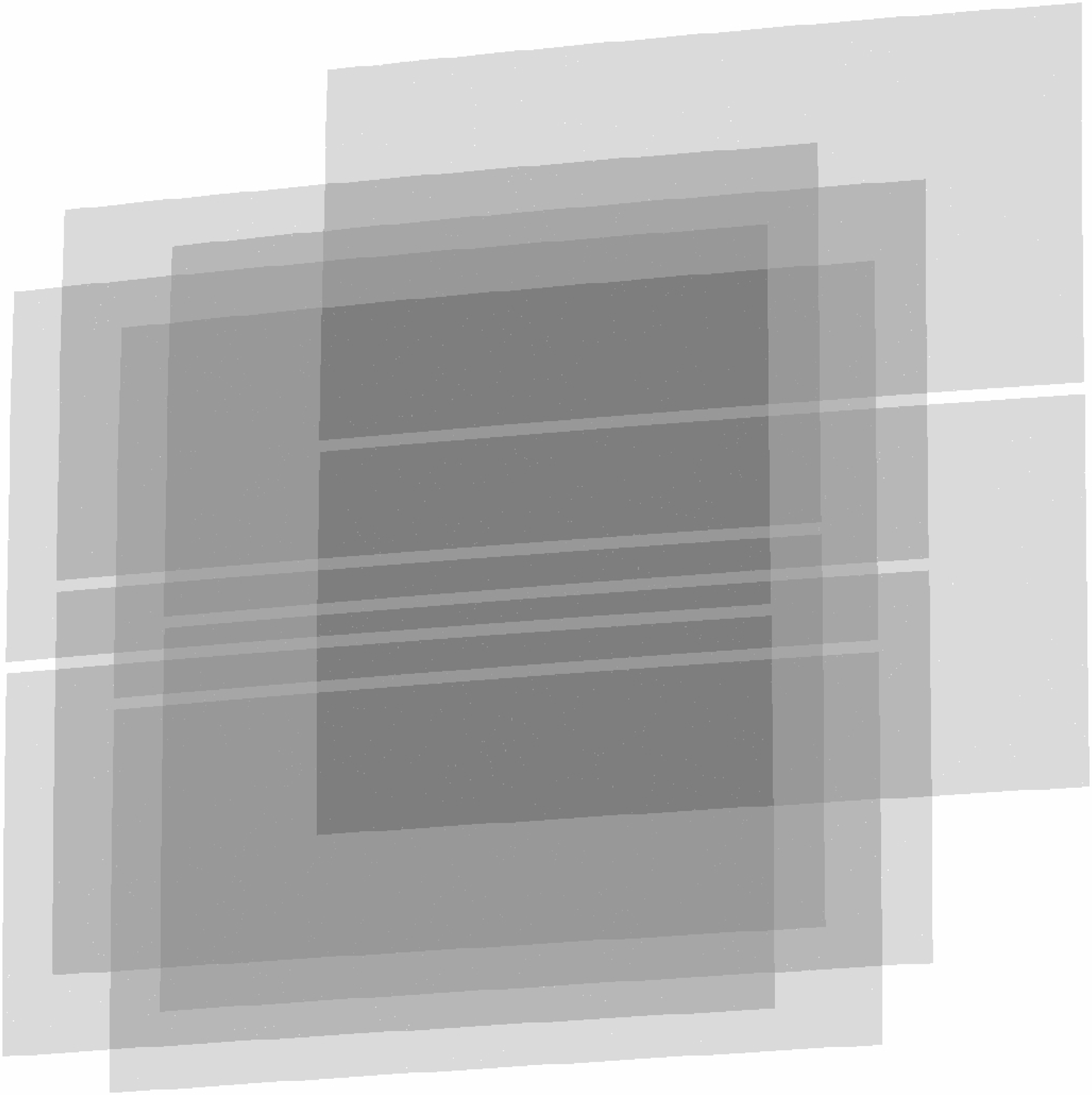}
\includegraphics[width=55mm,height=55mm]{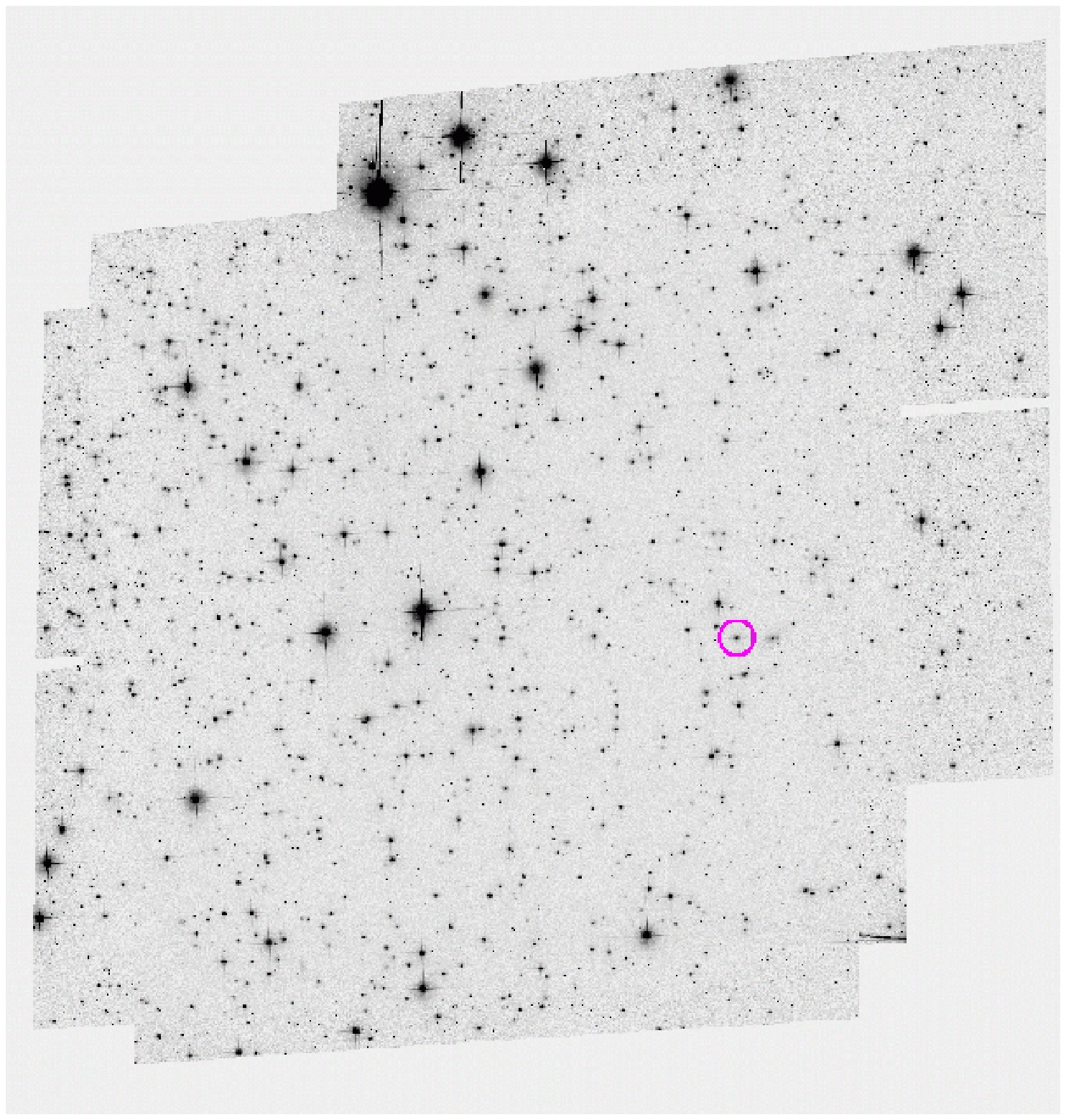}
\includegraphics[width=55mm,height=55mm]{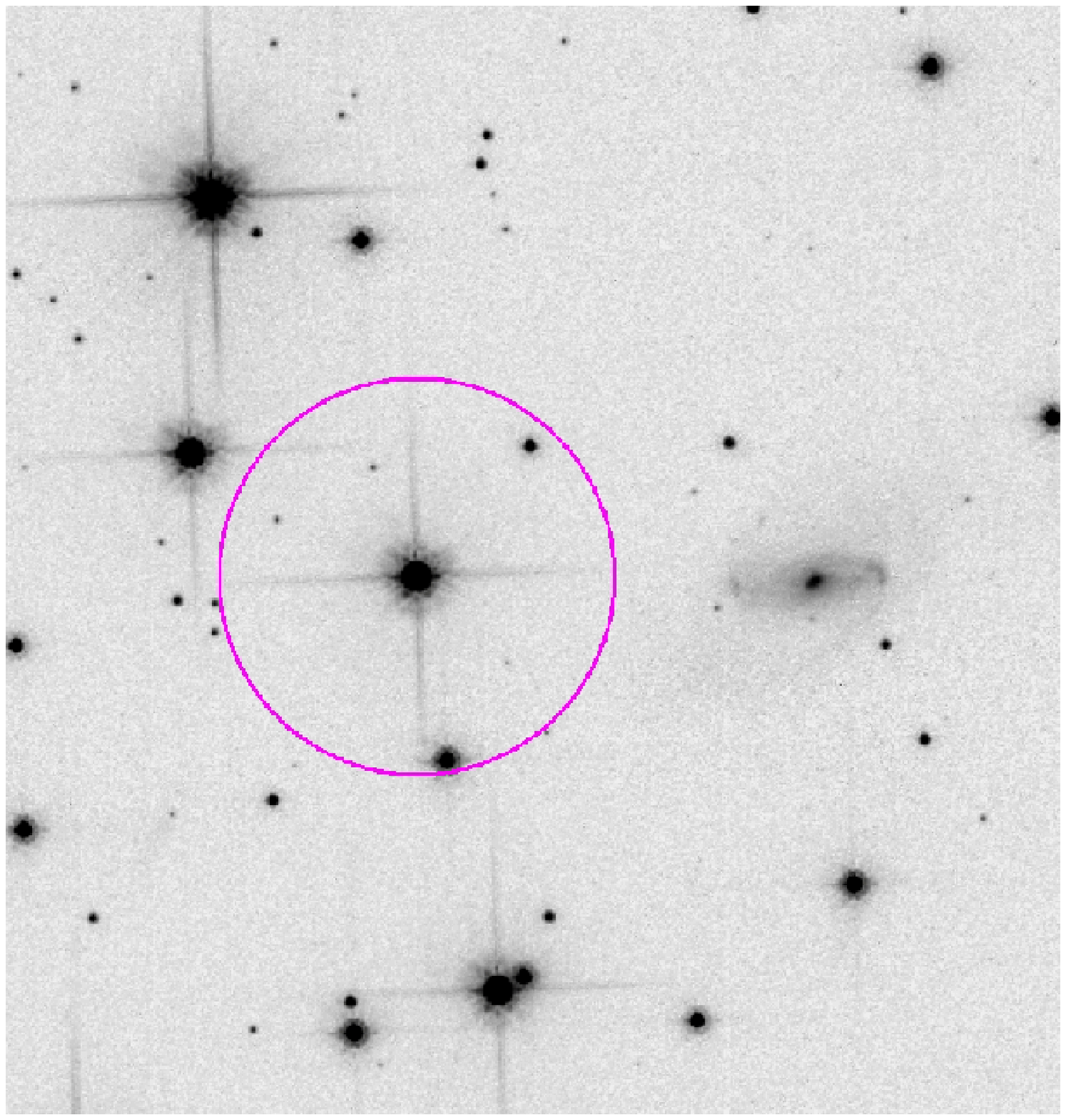}
\caption{
\textit{(Left:)} depth of coverage map for the five ACS/WFC/F814W images 
collected within epoch\,01. 
\textit{(Middle:)} resulting stack image. 
\textit{(Right:)} a $\sim$$27^{\prime\prime}$$\times$$27^{\prime\prime}$ 
zoom-in around the star marked with a circle in the previous panel. For
reference, the circle has a 200-pixels radius (in the ACS/WFC stack the
pixel scale is $\sim$25 mas).   
}
\label{acswfc}
\end{figure*}

\begin{figure*}
\includegraphics[width=150mm,height=150mm]{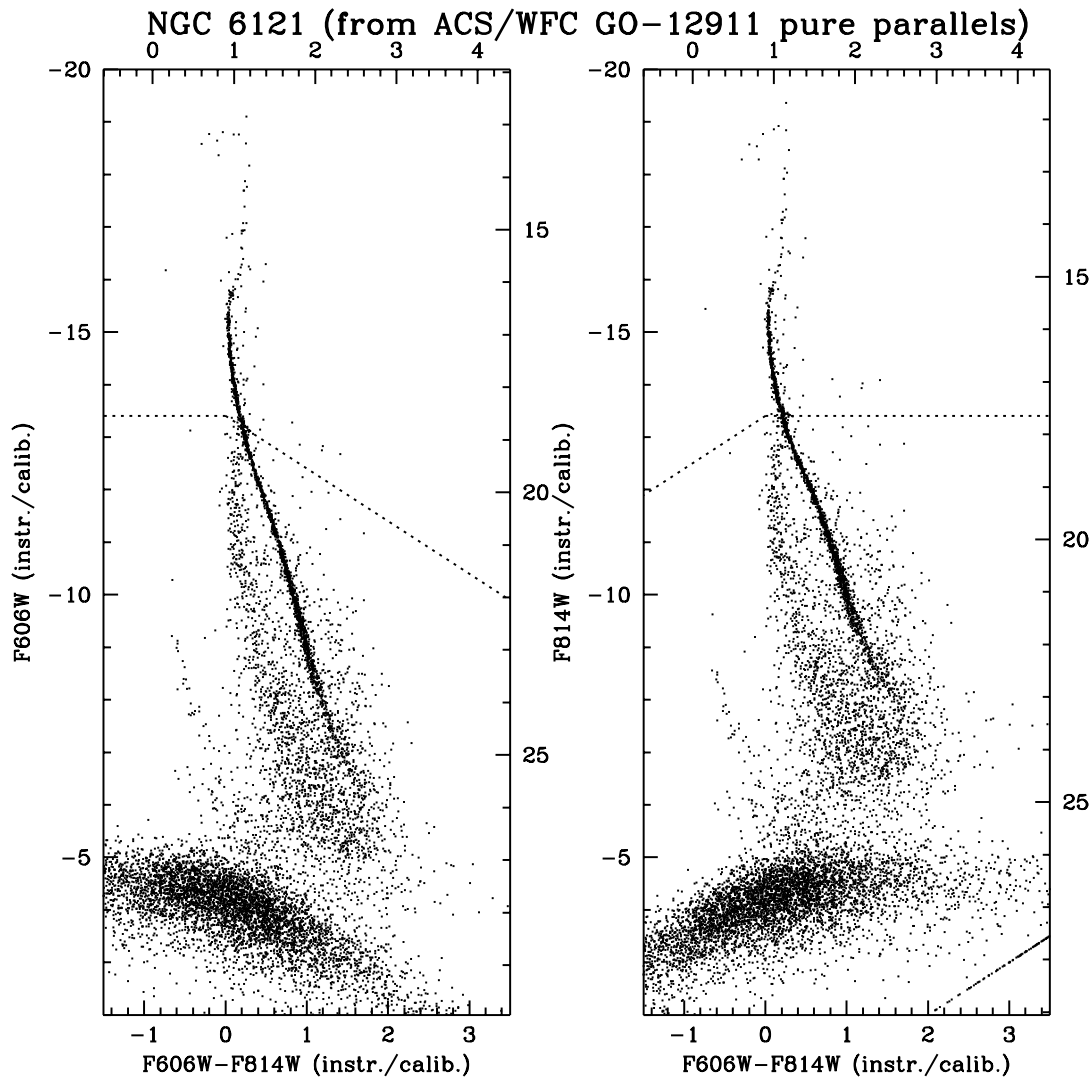}
\caption{
Color-magnitude diagrams for the coordinated parallel observations
with ACS/WFC using the 5 images for each filter.  The distributions
of dots fainter than instrumental magnitudes of $-$5 and centered at 0
in color, indicate the floor-level of white noise.  }
\label{CMDsACS}
\end{figure*}

In Fig.~\ref{acswfc}, we show on the left the depth of coverage for
the ACS/WFC parallel images of epoch\,01 in filter F814W, in the
middle the corresponding stack image, and on the right a close-up of a
representative $\sim$27$\times$27 arcsec$^2$ region.
Figure~\ref{CMDsACS} shows the color-magnitude diagrams obtained for
stars in this ACS parallel field.  The dotted lines mark the on-set of
saturation. In the plots are given the instrumental colors and
magnitudes (bottom and left axes), and the calibrated ones (top and
right axes).

\subsubsection{WFC3/UVIS}
\label{Swfc3uvis}

The WFC3 team has recently found (Baggett \& Anderson 2012) that a
small amount of post-flash ($\sim$12\,e$^{-}$) can prevent a
significant amount of CTE losses.\footnote{
See the document
\textsf{http://www.stsci.edu\-/hst\-/wfc3\-/ins\_performance\-/CTE\-/CTE\_White\_Paper.pdf}.
} 
In planning the observations, we examined the background in archival
F467M images taken of the M\,4 core for GO-12193 (PI: Lee) and found
that we should expect about two electrons per pixel in a typical
350\,s exposure.  To get the recommended background of 10-12\,e$^{-}$,
we decided to add 8\,e$^{-}$ of post-flash to each F467M exposure.
The background achieved in the first visit is about 15\,e$^{-}$, a bit
higher than expected, but the mitigation should still be quite good.

As mentioned in Sect.~\ref{ctem}, the WFC3 team is also currently
working to develop a pixel-based CTE-restoration capability similar to
what has been done for ACS/WFC.  This was not available at the time of
the first epoch, so this preliminary analysis has been done on
uncorrected images.

We measured star positions and fluxes on the WFC3/UVIS \textsf{\_flt}
images with a routine that is similar to that described in AK06 for
the WFC/ACS.  We used ``library'' PSFs that were extracted from F467M
observations of $\omega$\,Centauri that accounted for the static
spatial variability.  We also constructed a distortion solution for
F467M from the same data set.  (As more data become available for this
field, this program will allow a definitive distortion solution and
average PSF for F467M.)  These positions and fluxes were corrected for
distortion and used to relate each exposure to the master frame.

\begin{figure*}
\includegraphics[width=55mm,height=55mm]{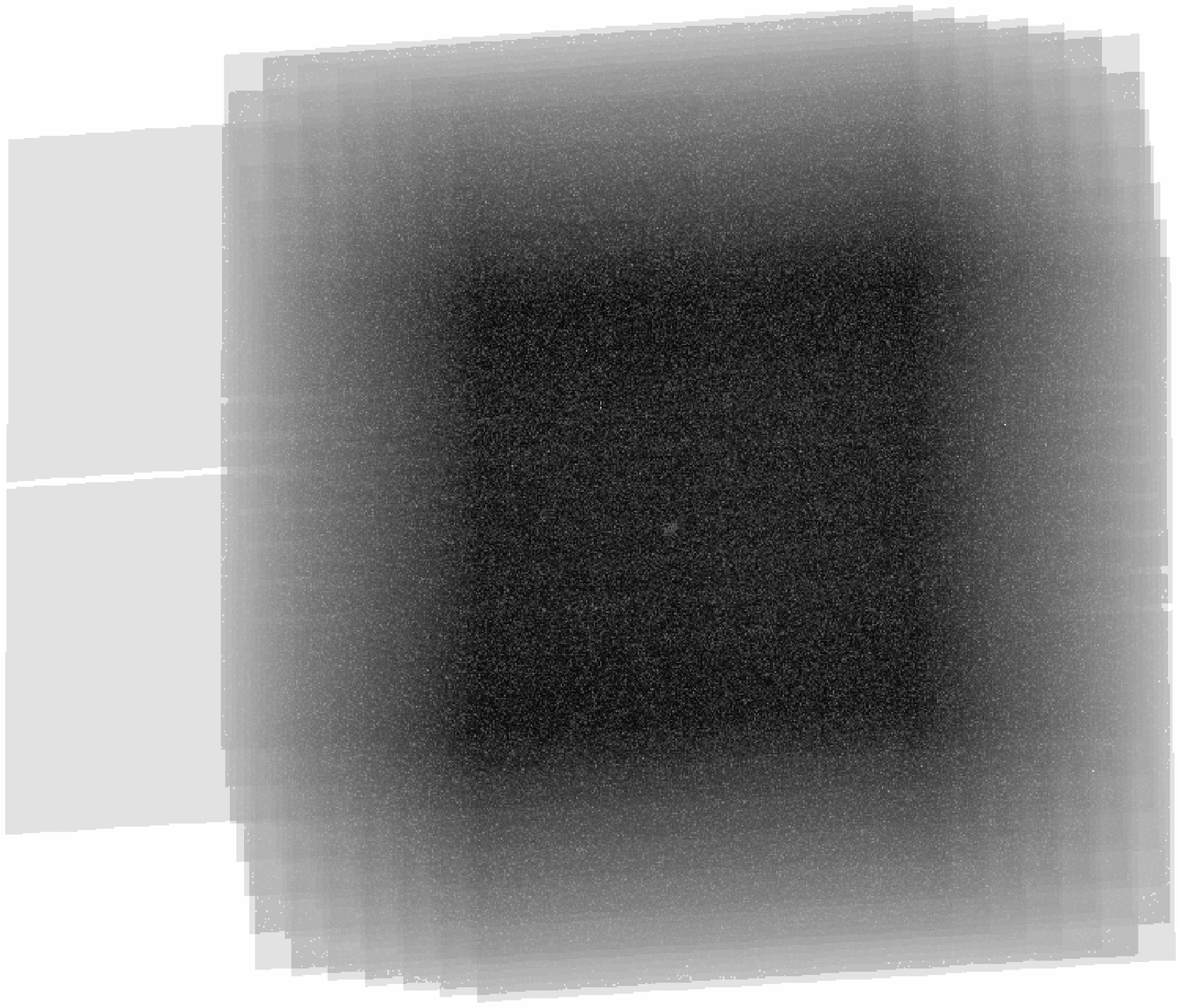}
\includegraphics[width=55mm,height=55mm]{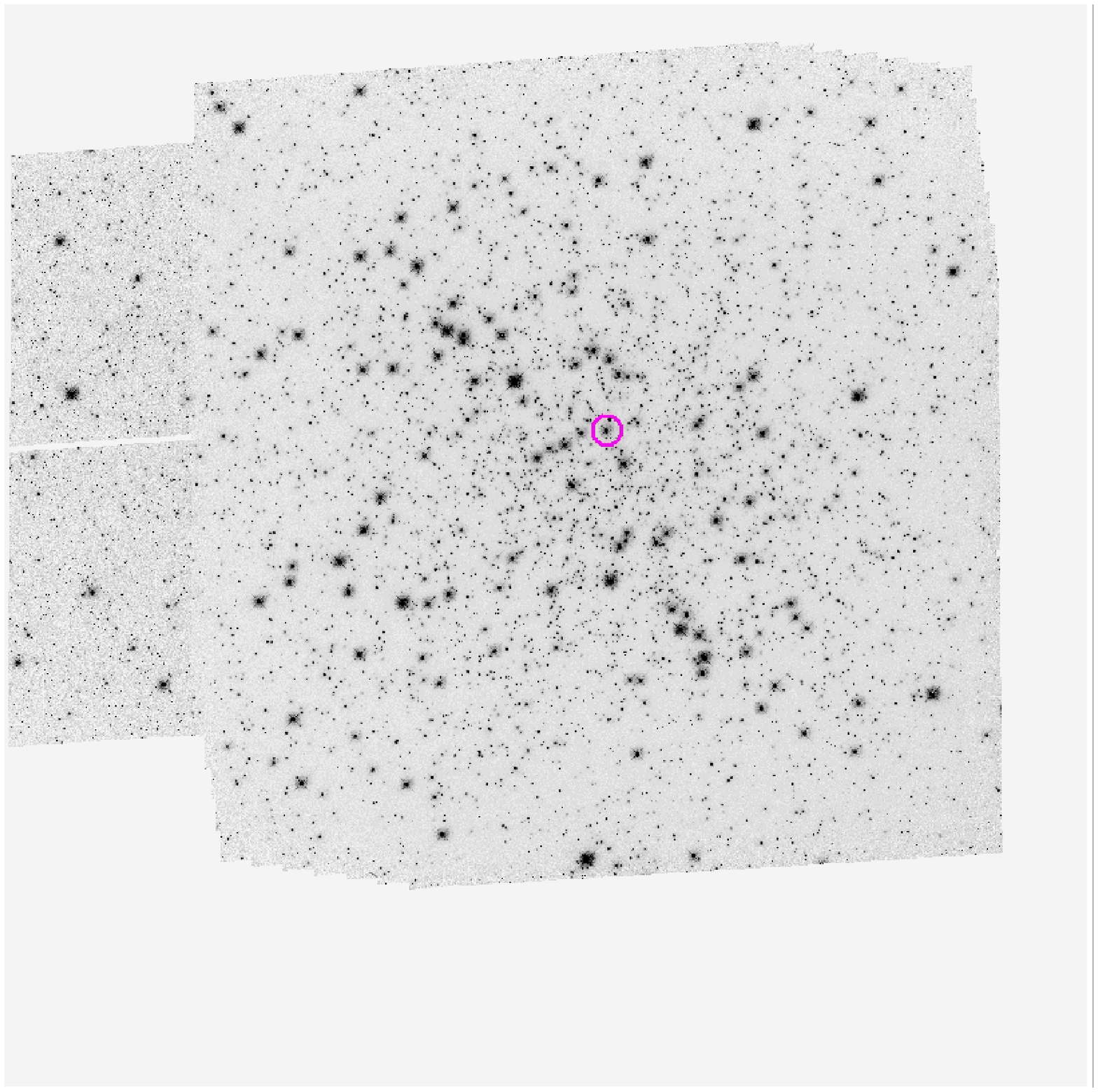}
\includegraphics[width=55mm,height=55mm]{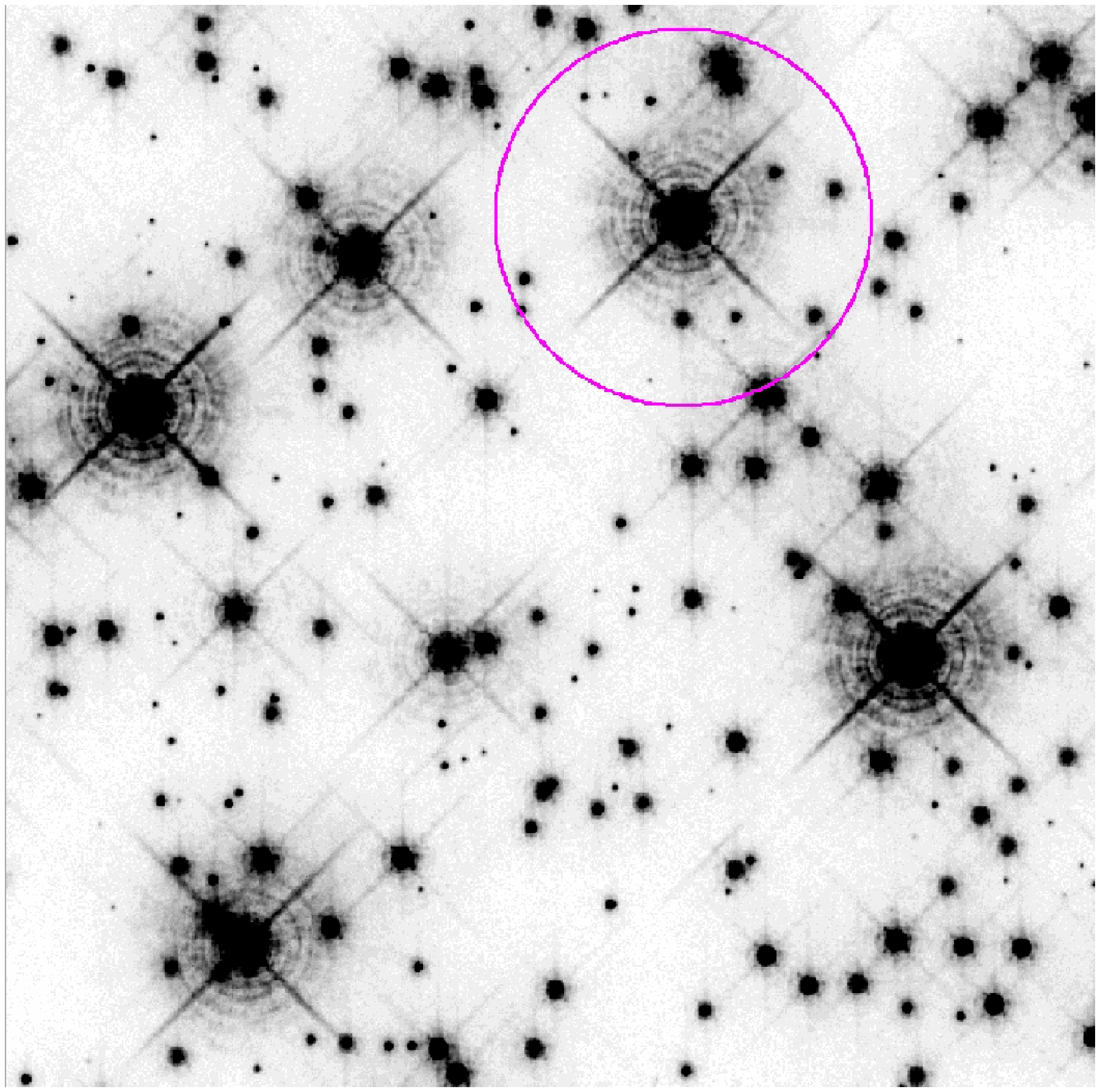}
\caption{
\textit{(Left:)} depth of coverage map for the 50 
WFC3/UVIS/F467M deep astrometric images of epoch\,01. 
\textit{(Middle:)} resulting stack image. 
\textit{(Right:)} a $20^{\prime\prime}$$\times$$20^{\prime\prime}$
zoom-in at the center of M\,4. For reference a circle with a
200-pixels radius is shown (WFC3/UVIS stack has a pixel scale of
$\sim$20 mas).  Note that diffraction spikes will have different
orientations at the different epochs.
}
\label{wfc3uvis}
\end{figure*}

\begin{figure*}
\includegraphics[width=150mm,height=150mm]{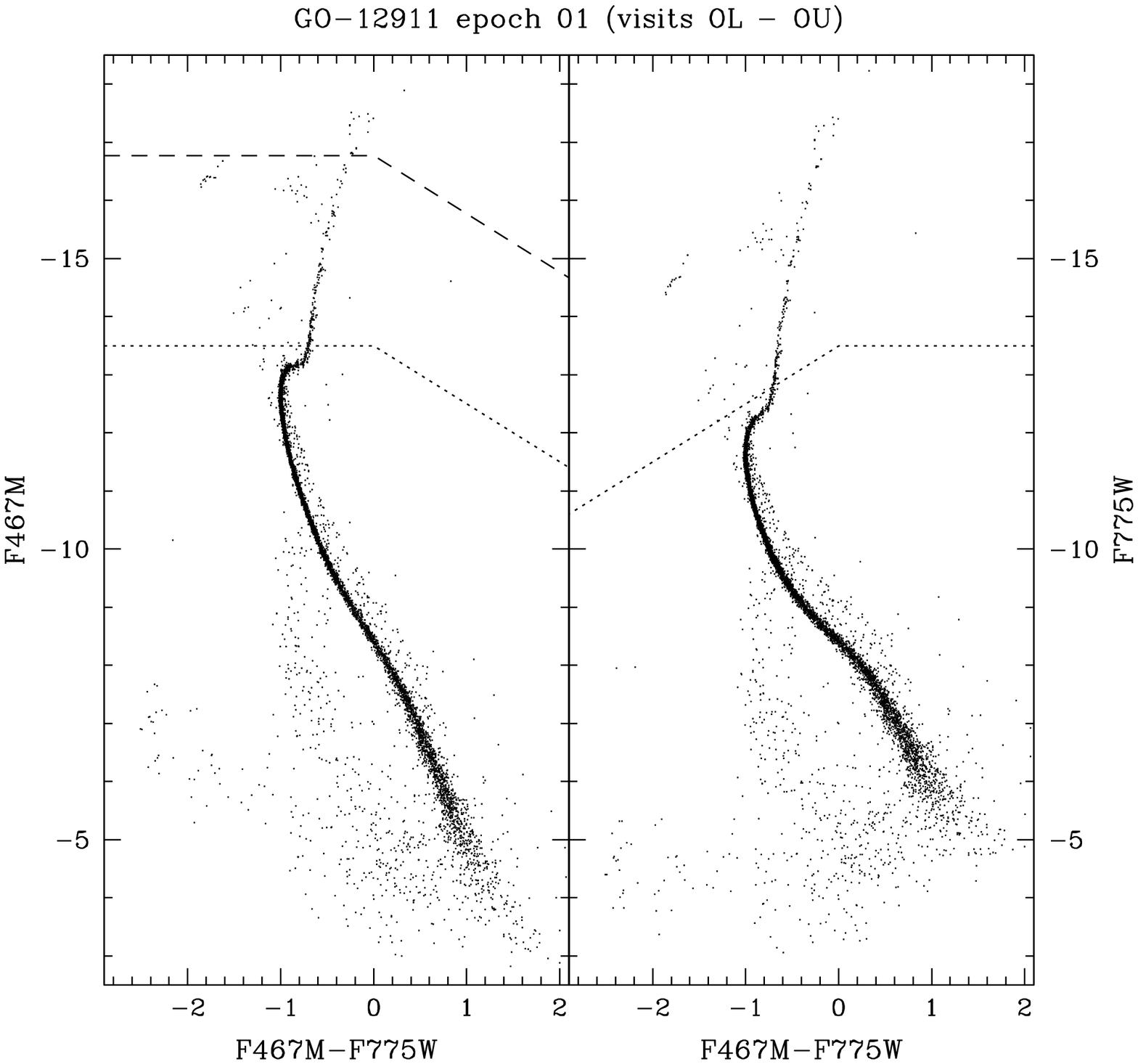}
\caption{
Color-magnitude diagrams of sources in the WFC3/UVIS field, obtained
by combining the photometry from the 51 F467M images and the 9
WFC3/UVIS/F775W images. Magnitudes are instrumental, and the dotted
lines indicate the on-set of saturation. Dashed line marks on-set of
saturation for the 20\,s short exposure in F467M.
}
\label{CMDsWFC3}
\end{figure*}

We then used a software program that is similar to that developed by
Anderson et al.\ (2008) for the \textit{ACS Survey of Galactic
  Globular Clusters} (GO-10775, Sarajedini et al.\ 2007) treasury
project to go through the field in 120$\times$120-pixel tiles to
iteratively find and measure stars simultaneously in all 50
contributing WFC3/UVIS/F467M 390\,s exposures.  Figure~\ref{wfc3uvis}
shows the depth of coverage for the WFC3/UVIS observations, an image
of the field, and a close-up of a representative 20$\times$20
arcsecond region. Figure~\ref{CMDsWFC3} shows the color-magnitude
diagrams obtained from this initial finding run in the WFC3/UVIS core
field.  The dotted lines mark the saturation level (the dashed line
marks the on-set of saturation for 20\,s short exposures in F467M).


%
\subsection{Preliminary astrometric precision} 
\label{pap}

The motivation for this program came from the finding in Bellini et
al.\ (2011) that stars can be measured with a precision of $\sim$0.35
mas in each coordinate in a single exposure.  We show below that even
our preliminary reduction of the first epoch already achieves this
goal.

\begin{figure*}
\includegraphics[width=170mm,height=85mm]{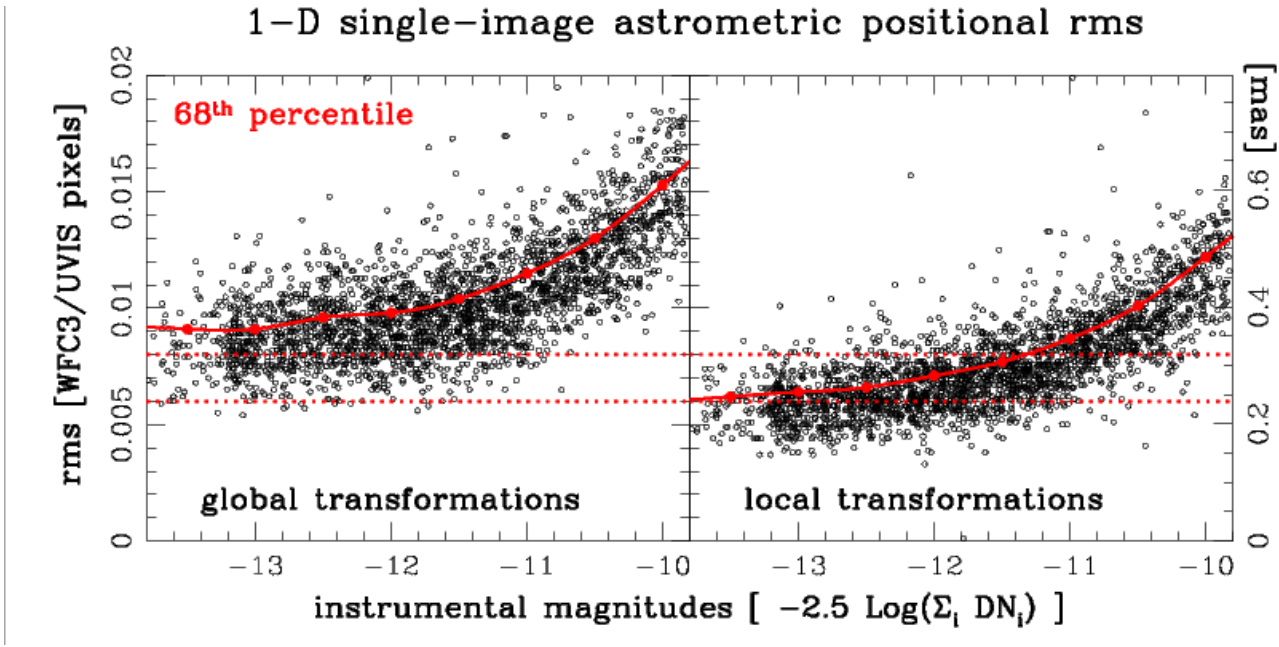}
\caption{
\textit{(Left):\/} 
Each small black circle is the root mean square (rms) of the
$x$-positions of unsaturated stars measured in at least 38 (and up to
50) individual WFC3/UVIS/F467M images once transformed ---with the
most general linear \textit{global} transformation--- into the adopted
reference frame, and plotted as function of the instrumental
magnitude. Every half-magnitude bin, a red dot indicates the positions
of the 68$^{\rm th}$ percentiles of the distribution, $\sigma$.  A
solid line (in red) connect these points.  For reference two
horizontal and dotted lines (also in red) mark the 0.006 and 0.008
WFC3/UVIS-pixel levels (i.e.\ 0.24 mas and 0.32 mas).
\textit{(Right):\/} 
Same plot for same stars, but this time using \textit{local}
transformations.  The positioning random error for the brightest stars
is now $\sim$0.25 mas.  The reduction is still preliminary (see text)
but 50 dithered images with this precision should be able to lower the
error on the mean $\overline{\sigma}$ to below the required detection
limits of $\overline{\sigma} \simeq 50 \mu$as.  Units of WFC3/UVIS
pixels are given on the left, while the ordinate on the right reports
units of milli-arcsec (mas).
}
\label{rmsX}
\end{figure*}

To determine astrometric uncertainties, we measured each star in each
exposure, corrected its position for distortion, then used a general
6-parameter global linear transformation to estimate its position in
the reference frame.  The scatter of these many estimates gives us an
indication of the measurement uncertainty.  The left panel of
Fig.~\ref{rmsX} shows the 1-D rms among the measured positions.

We computed the 68th percentile of the distribution of these rms in
every half-magnitude bin (red filled circles). A solid line (in red)
through these points highlights the behavior of the rms as a function
of the magnitude (i.e.\ as a function of the S/N).  This shows that
even with a global transformation the positional errors for the
typical star just below saturation is on the order of $\sim$0.01
WFC3/UVIS pixels (i.e.\ $\sim$0.39 mas).  Two dotted lines mark the
0.008 and 0.006 pixel level (i.e.\ the 0.24 and 0.32 mas precision
level).

We find that these errors are further reduced when we use
\textit{local} transformations from each frame into the reference
frame.  Local transformations absorb most of the residuals in the
geometric distortion due to the imperfection on the PSF models,
detector features, particular status of the focus, and of the optical
alignment in general (see Bedin et al.\ 2003 or Anderson et al.\ 2006
for more details).  In computing local transformations we employed the
101 stars closest to the target, within 500 WFC3/UVIS pixels, with an
instrumental magnitude brighter than $-10$. The right panel of
Fig.~\ref{rmsX} shows the 1-D precision for this local transformation
approach. This plot demonstrates that a precision of $\sim$0.25 mas
can be reached for the brightest unsaturated objects (i.e., almost
30\% better than our nominal requirement).

We want to emphasize that these results are still preliminary for the
following reasons:
\begin{enumerate}  
\item in this data reduction, no pixel-based mitigation of CTE problems was performed, 
   neither in $x$ nor in $y$ (serial and parallel readings);  
\item the PSFs were taken from a library; they are spatially variable 
   but not tailored to the particular focus-state of the telescope during the 
   given individual exposure;
\item the geometric distortion correction for filter F467M is still crude; 
   in the end we will expect to have an almost perfect
   characterization of the geometric distortion in this filter,
   significantly better than what was done for the 10 filters in Bellini
   et al.\ (2011), thanks to the huge ---unprecedented--- amount of
   data available.
\end{enumerate}  
We also plan to attempt a modeling of the perturbation to be applied
to both the average PSFs and the average geometric distortion
correction, as a function of the \hst\/ position telemetry, angle
between Sun and telescope tube, temperature, etc, using de-correlating
techniques.
Nevertheless, we note that, in spite of the crudeness with which it was
derived, the above-described 1-D precision of $\sim$0.3 mas, for a
well-exposed star, \textit{already} meets our project requirements.\\ 

%
%

It will be extremely easy to separate background stars from cluster
stars by means of the proper motions.  The field-cluster separation is
hundreds of times larger than the astrometric wobbles of binaries we
propose to measure here.
Field-objects mainly belong to the outskirts of the Galactic Bulge,
which is in the background of M\,4.  To demonstrate this PM separation,
we downloaded from \hst\,'s archive images collected under program
GO-10775 (PI: Sarajedini) on May 3rd, 2006 (Sarajedini et al.\ 2007),
and use them as a previous epoch.  We reduced \textsf{\_flc} ACS/WFC
images in filters F606W and F814W, 4 per filter of 25-30 s, and
extracted fluxes and positions.

\begin{figure*}
\includegraphics[width=170mm,height=132mm]{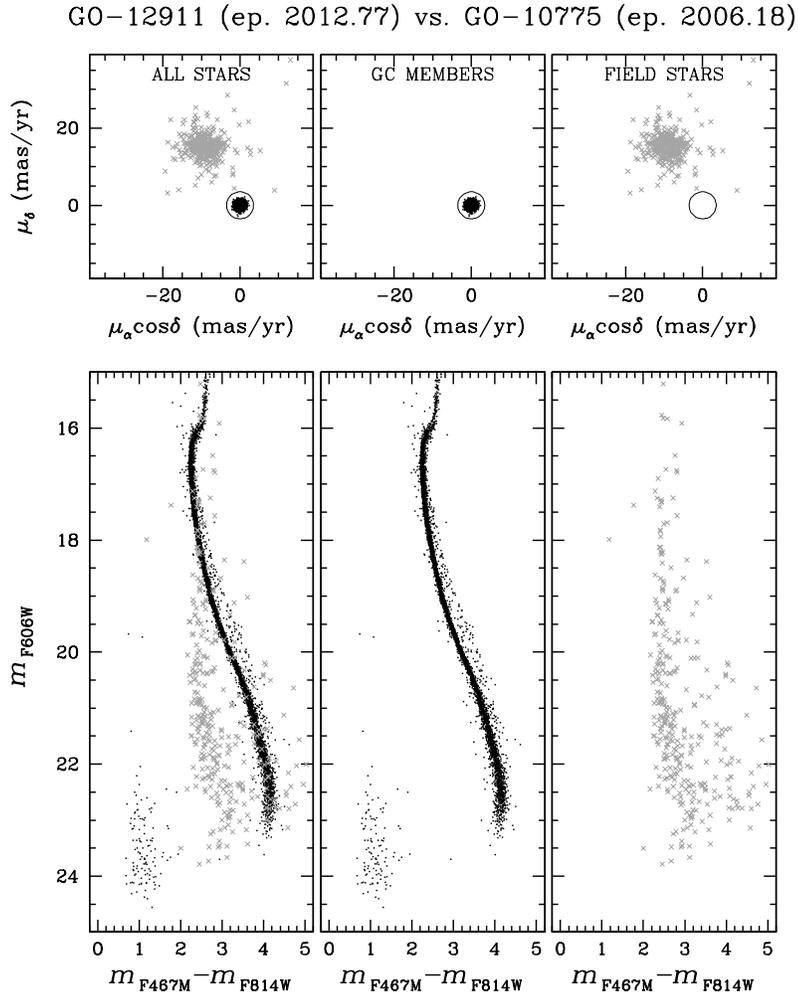}
\caption{
\textit{Primary field in the core of M\,4:\/}
\textit{(Top):\/} 
Proper-motion diagrams (in mas yr$^{-1}$ units) obtained comparing
displacements between average positions measured in \textit{epoch\,01}
and those measured in the archival ACS/WFC images collected in 2006
under program GO-10775.
\textit{(Bottom):\/} 
Corresponding calibrated CMDs for objects in the above proper-motion diagrams. 
\textit{(Left):\/} 
The entire sample of all common objects to the two data-sets.  
\textit{(Middle):\/} 
Objects assumed to be members of the GC, i.e.\ with proper motions smaller than 3.5 mas yr$^{-1}$
(criterion highlighted by circles in the top panels). 
\textit{(Right):\/} 
Objects with proper motions greater than 3.5 mas yr$^{-1}$, and
considered field objects. 
In all panels, members are indicated with dots and field objects with crosses. 
}
\label{pms}
\end{figure*}

In Figure~\ref{pms}, the top panels show the proper-motion diagrams for the 
common objects to the two data sets.  In the bottom, the corresponding 
calibrated color-magnitude diagrams are shown.  From left to right, the 
plots refer to the entire sample, the sample with cluster motion (arbitrarily 
defined as proper motions smaller than 3.5 mas yr$^{-1}$), and the objects 
with field motions, i.e.\ with proper motions larger than that.   

\begin{figure*}
\includegraphics[width=85mm,height=85mm]{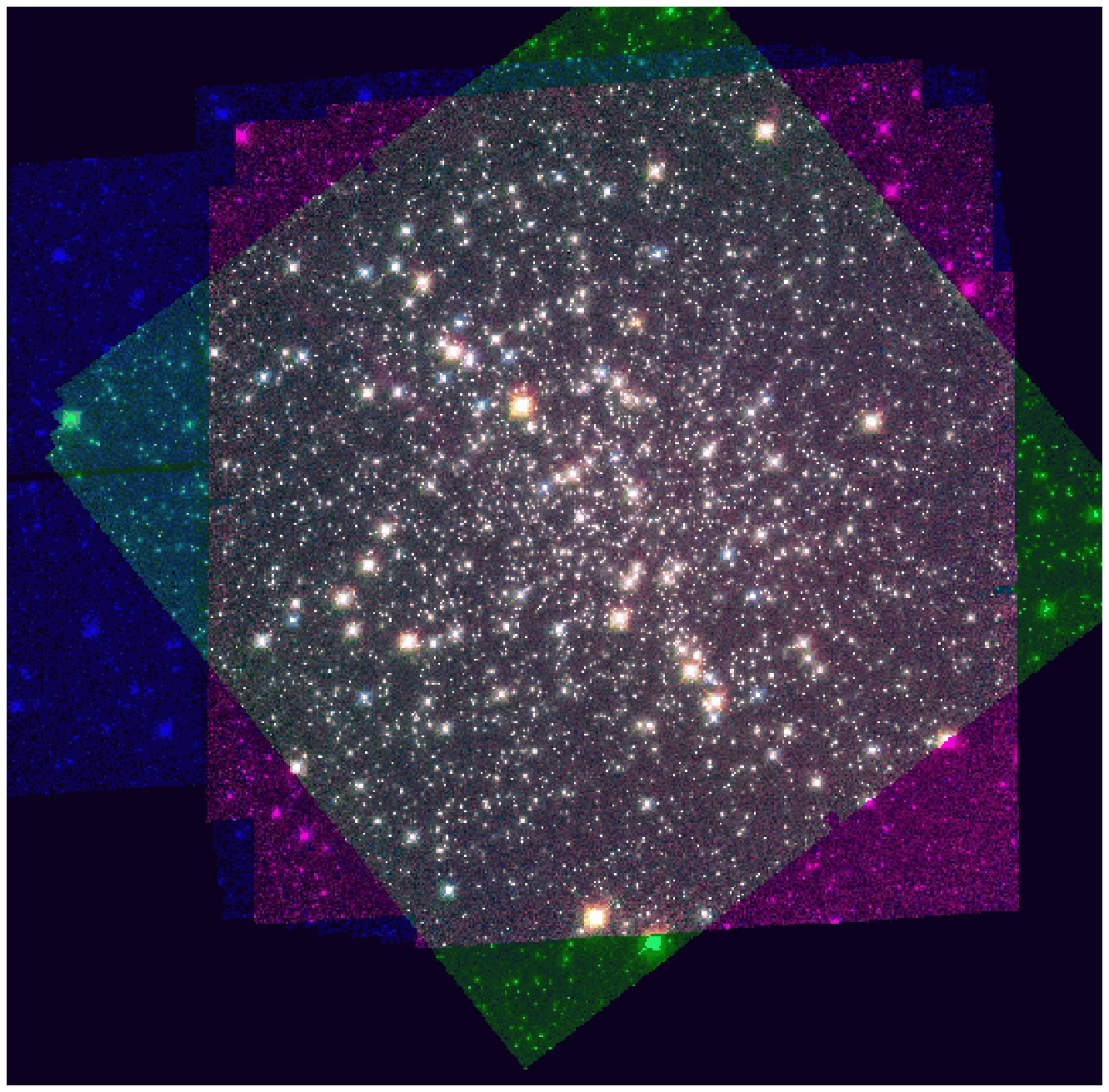}
\includegraphics[width=85mm,height=85mm]{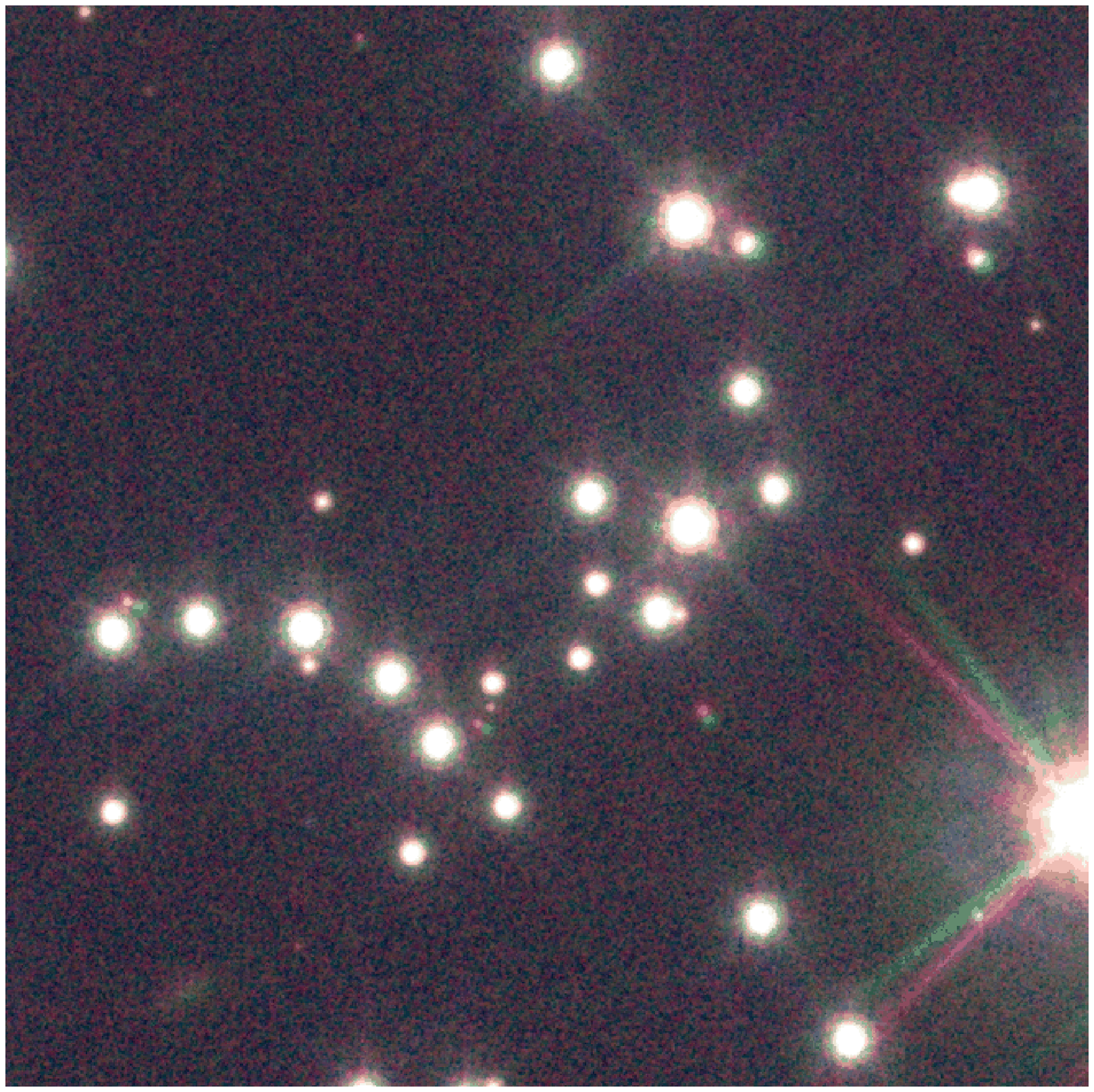}
\caption{
\textit{(Left\/):} 
Three-color image of the core of M\,4, where for the blue- and
red-channels we used the WFC3/UVIS/F467M and F775W GO-12911 images,
and for the green-channel we used ACS/WFC/F606W images from 
GO-10775 collected $\sim$6 years earlier.
\textit{(Right\/):}
Close-up of a region in the same image of $\sim$$13^{\prime\prime}$$\times$$13^{\prime\prime}$. 
Note that, as the transformation is computed with respect
to cluster members, which define the zero of the motion, the field
objects appear as trailed rainbows.
}
\label{rgb}
\end{figure*}

Figure~\ref{rgb} shows a three-color stacked image of the central
field\footnote{ 
The image uses F467M for the blue channel, F775W for the red channel,
and F606W (from GO-10775) for the green channel.
}.  
Since the cluster stars were used as the basis for the transformations,
they have no motion in this frame.  But field objects appear as
rainbow trails, with an offset between their green and red+blue light.
 

%
\subsection{Preliminary photometric precision} 
\begin{figure*}
\includegraphics[width=170mm,height=85mm]{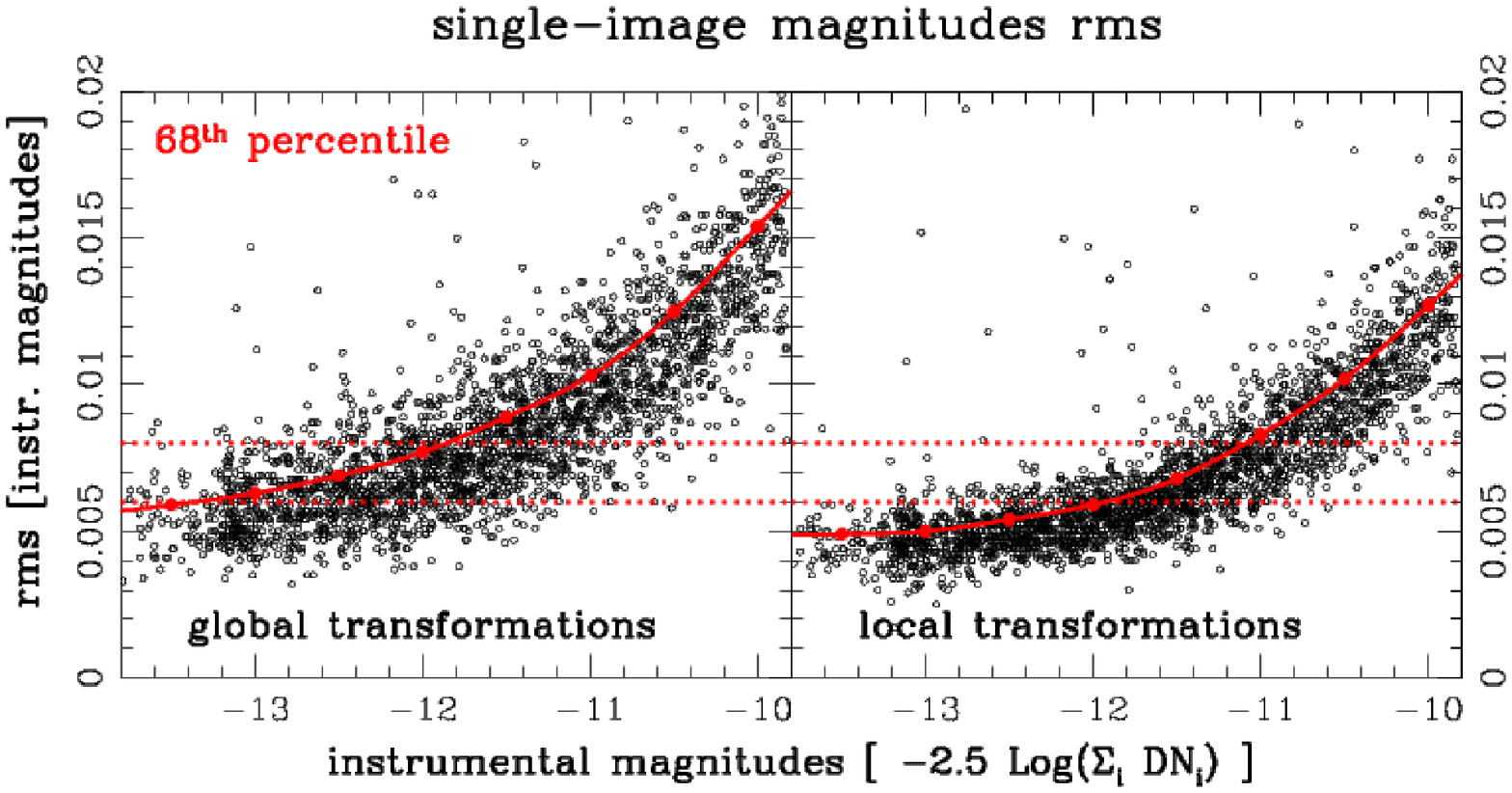}
\caption{
As in Fig.~\ref{rmsX} but for magnitudes instead of $x$-positions.
Ordinates axes are all in units of magnitudes, and the two dotted
horizontal lines mark the level of a 6 and 8 milli-mag precision.
}
\label{rmsM}
\end{figure*}

As noted before, the observing strategy of GO-12911 is not meant to be
optimal for differential photometry, as the adopted large dithers leave 
our photometry at the mercy of the accuracies of the flat-fielding
(both L-flats and P-flats), and precise photometry is also affected by
the locations of artifacts and manufacturing features, which vary their
positions with the different dithers and orientations. 
Nevertheless, the photometric accuracy is still quite good, even for
our preliminary reduction.
Figure~\ref{rmsM} is the photometric analog of Fig.~\ref{rmsX} for
positions, where instead of 1-D  rms of $x$-positions we show the
rms in magnitude.  Indeed, it is possible to apply local corrections also
to fluxes to remove systematic trends.  We used the same network of
local stars as in the previous figure.  Figure~\ref{rmsM} demonstrates 
that for the brightest stars we are able to achieve a photometric accuracy 
of $\sim$5 milli-magnitudes, i.e., still significantly worse than 
shot-noise ($\sim$3 milli-mag).  Using techniques similar to those 
presented in Nascimbeni et al.\ (2012),  we expect our final photometry 
to further approach the theoretical limit.


\section{Planned upcoming works}
\label{papers}

The final data-set of program GO-12911 will consist of 840 \hst\/ images 
(720 WFC3/UVIS primary and 120 ACS/WFC parallels) and will have many 
\textit{unique} characteristics that were not available in any of the 
previous \hst\/ programs (see Sect.\ref{cmp}).  
The target and cadence of this data set enable a large number of
studies, and our team has identified our top priorities.  
Our focus will be on:
a search for binaries with massive and dark companions (the main
project); a search for exo-planets; a search for stellar variability
and X-ray$\setminus$optical counterparts; the measurement of the
cluster absolute proper motion and of the cluster parallax; the
measurement of internal motions of stars in the cluster core; the
search for an intermediate-mass black hole; and the study of the
multiple stellar populations in the cluster.  All of these
sub-projects together will produce improved constraints for our
dynamical modelling of the present-day M\,4 and of its dynamical
evolution history (Sect.~\ref{dyn_mod}).
This unique data-set will also allow us to perform a number of
technical studies and instrument calibrations, which we will make
publicly available (Sect.~\ref{b-p}).
One of the major outcomes of this project will be a catalog of sources
in the center of M\,4 of great richness (Sect.~\ref{cat}).

In the following sections, we describe in more detail the particular
sub-projects (which we also list with the time-frame in
Table~\ref{tab03}) that our team will be focusing on.

\subsection{Binaries with massive dark companions}
\label{bwmdc}

The fundamental goal of the project is to detect the fraction of
binaries with a massive and dark component, through the astrometric
wobbles induced on their luminous companions, within the GC M\,4.
The motivation for this investigation and the currently-expected
properties of these binaries were given in Sections 2 to 4, while
Sections 5 and 6 described how this goal has driven our observing
strategy.
The detected wobbles will provide periods and semi-major axes,  which
in turn can provide the absolute masses, and mass-ratios $q$ of these
binaries.  These will be used to constrain our dynamical model
(Sect.~\ref{dyn_mod}) and infer fundamental properties of the
primordial binaries, a basic ingredient for tracing the dynamical
evolution of clusters.
In this section we briefly describe the analyses of reduced data and
publication strategy for this task.


In order to analyze our data set, we need to find the best average
(spatially variable) PSF and the best average distortion solution
(these are related to each other).  This will allow us to find the
best position and proper motion for each star.  With that, we can
predict the position of each star in each exposure.  To measure a
given star, we will first find its nearest bright isolated neighbors
within $\sim$500 pixels (as motivated in Sect.~\ref{nee} and shown in
Sect.~\ref{pap} for epoch\,01).  We will use the predicted positions and
fluxes for these neighbors to extract a local PSF from them (actually,
a local perturbation on the global PSF for that location). Then, we
will use this local PSF to measure an empirical position for the
target star.  This will give us a position that can be directly
compared against the predicted position (based on the linear proper
motions). 
These residuals can be used to either improve the average 
positions/motions or to look for wobble.

We will necessarily have to wait for all epochs to be collected.  The
binary-system barycentric motions need to be solved for and
disentangled from their wobbles and residuals using a periodogram-type
analysis to detect significant and periodic deviations from the linear
motions.  Most of the effort will go into identifying sources of
systematic errors and in correcting for them.  Artificial-star
simulations will be an important aspect of this effort.

Some photometric modulation of the light curve, phased with the
astrometric wobbling signature, might also be present as a result of
ellipsoidal variations and synchronization of the spin and orbital
motion induced by the dark companion on the luminous member of the
binary system. These effects would provide us with additional
confirmation of the detected wobbles.


\subsection{Search for variable stars}
\label{var}
Our survey will generate $\sim$600 high-quality F467M images of the
core of M\,4, split into 12 epochs (each one covering $\sim$15
nearly-consecutive hours) and spanning a year. We can also take
advantage of some color information extracted from the F775W images
taken at the beginning of each orbit ($\sim$120 frames overall).  We
will search for variability of every point-like source robustly
detected in a minimum number of frames. The photometric signal of a
variable star can span an extremely large range of amplitudes, shapes
and periodicities according to its class (eclipsing, pulsation,
rotation, cataclysmic variable, etc...). We will tune different
software tools in order to maximize the detection rate of each
variable class, including weighted and multi-harmonic periodograms
(Schwarzenberg-Czerny 1999; Zechmeister \& Kurster 2009).

Eclipsing binaries (EBs) are expected to dominate the sample of
photometrically detected variables among cluster members. This is a
fourth, complementary, approach to finding binaries in M\,4, after
detection through radial velocity (RV) variations, photometric offset
in the CMD (high $q$ MS+MS binaries), and astrometric wobbles. The
average photometric precision ($\sigma \lesssim 5$ mmag) on the
brighter part of our sample and the time-sampling, are such that we
expect a non-negligible efficiency up to long orbital periods
($P\lesssim 15$ days).
M\,4 is intrinsically much richer in binaries than other well-studied
GCs such as NGC\,6397 and 47\,Tuc (Milone et al.\,2012a).  M\,4 is also 
dynamically old and we expect most binaries to have sunk into its
central region.  By rescaling the yield of EBs estimated by Nascimbeni
et al.\,(2012) in NGC\,6397, we expect a few tens of detections from
our data set. Some of them will be SB2 detached systems bright enough to
be suitable for a ground-based RV follow-up on 8-10\,m class telescopes.
For these, an accurate estimate of the masses and radii of
the binary components will be feasible, as further explained
in Sect.~\ref{debs}. 

Single rotational variables, whose periodicity is dominated by the
intrinsic rotational period of the star are another possible outcome
of this search. Gyrochronology tells us that stars in old systems such
as GCs should have stars with rotational periods of several months,
and photometric amplitudes of a few mmags (Chaname \& Ramirez
2012).\footnote{
Binary rotation variables have much shorter periods, 
as the stars are tidally locked. 
} 
We will be in the position to  measure some of these periods and,
possibly, for the first time to define  a gyrochronological sequence
for a GC. 

We will also search for variability on the ACS/WFC parallel fields.  A
large fraction of these fields (over 50\%) will be repeated during our
survey (from 5 to $\sim$30 times per filter), allowing us to detect periodic
variables in a much less crowded region of the outskirts of M\,4 (see
right panel in Fig.~\ref{obs}). This investigation will provide us
with important information on the radial distribution of the different
classes of variables.
However, we expect few binaries there, as 
by examining a 120 orbit WFPC2 integration in the outskirts of
M\,4, Ferdman et al.\ (2004) found a variability fraction of just
0.05\% for a deep field located at six core radii from the
cluster center.

\subsection{Search for transiting exo-planets}

Planets in star clusters offer a unique opportunity to probe how the
chemical and dynamical environment affects the formation and evolution
of planets.  To date, searches for planetary transits carried out in
GCs have yielded no confirmed detections (see Zhou et al.\ 2012 for a
review).  In the two most significant cases (47\,Tuc and $\omega$ Cen)
these results imply a lower occurrence of giant planets compared with
field stars in the solar neighbourhood (Gilliland et al.~2000;
Weldrake et al.~2008).  Still, it is not clear how and why the cluster
chemical or dynamical environment inhibits planetary formation or
survival, and whether smaller planets such as Neptunians are affected
in the same way as gaseous giants.

Some of us previously performed a search for transiting planets and
variables in a deep stellar field of NGC\,6397 imaged by \hst-ACS/WFC
over 126 orbits (Nascimbeni et al.~2012).  The light curves were
corrected for systematic trends and inspected with several tools,
including a box-fitting least-square algorithm (BLS, Kov\'acs et
al.\ 2002) to perform transit-finding in the BLS periodogram.
These data were not optimized for a transit search, and no significant
conclusion was drawn on the planet occurrence in NGC\,6397 due to poor
statistics and temporal sampling. However, we did demonstrate that most
NGC\,6397 MS members (98\%) are photometrically very stable, as they
show an intrinsic jitter below $\sim$2 mmag on the time scale of a
typical transit (few hours), and we expect the intrinsic jitter to be
of the same order of magnitude for M\,4.

We have run ``inject-and-recover'' simulations based on synthetic
light curves that have the same noise level and sampling times as the
12-epoch photometric series expected at the end of the program.  The
observing strategy is still far from optimal for transit searches,
since this program is designed for other purposes. Nevertheless, the
advantages of GO-12911 data over those used by Nascimbeni et
al.\ (2012) are evident (faster cadence, larger temporal coverage,
larger number of imaged cluster members).  We will achieve a
$\sim40\%$ efficiency in detecting transiting ``hot Jupiter'' planets
with orbital periods shorter than 4 days. Chances to detect a
Neptune-size planet are much lower (but not negligible: $\lesssim 5\%$
at $P<3$ days) due to their shallower transits.  On the other hand,
Neptunian planets are expected to be much more common than gaseous
giants, and --even more importantly-- their occurrence does not seem
to decrease at lower metallicities (Howard et al.~2012; Mayor et
al.~2011).  This will compensate the final expected yield of planets,
implying that in principle detections of Neptunian planets around the
MS stars of a GC could be feasible here for the first time.  At the
very least, we will be able to set an upper limit on planet frequency
in M\,4.
We end this section noting that at least one planetary-mass object is
already known to exist among member stars of M\,4 (Sigurdsson et
al.\ 2003).

\subsection{Absolute proper motions and annual parallax} 
\label{parallax}

To achieve the main goal of our project (i.e., to detect the wobbling
binaries) we will refer our proper-motion measurements to M\,4 member
stars, which are all at the same distance, so that we will not be
affected by either the absolute proper motion of the cluster or its
annual parallax.  At a distance of 1.7\,kpc, the annual parallax of
M\,4 is anything but negligible:\ $\sim$0.6 mas (i.e., 12 times larger
than our target accuracy for individual high S/N stars in a given
epoch).
So, in addition to the reflex of the absolute proper motion of the
cluster, we expect to see the few background stars, galaxies, and QSOs
(such as the one used by Bedin et al.\ 2003) to wobble by that much
with respect to the average of M\,4's stars.

Our epochs will map the annual parallax phases almost homogeneously,
and in combination with archival material (and also in combination
with our parallel fields), we will be able to derive one of the most
direct, accurate and model-independent estimates of the geometric
distance of M\,4, through the estimate of its absolute annual
parallax.  We will likely achieve an uncertainty below 5\% on the
distance, the exact value depending on the number of available
extra-galactic sources.

Most of the field objects are background stars belonging mainly to the
outskirts of the Galactic Bulge.
It should be straightforward to determine the relative parallax of
M\,4's stars (at $\sim$1.7\,kpc), with respect to the bulk of the
Bulge' stars in the field (at $\sim$8\,kpc, see Bedin et al.\ 2003).
For the few stars in the foreground of M\,4 in our main field, we will
derive exquisite relative annual parallaxes with respect to M\,4.

\subsection{Multiple stellar populations within M\,4}
\label{mPOPs}

Spectroscopic studies of bright RGB stars in M\,4, have shown
significant star-to-star light-element abundance variations (Norris
1981; Gratton et al.\ 1986; Brown \& Wallerstein 1992; Drake et
al.\ 1992; Smith et al.\ 2005), with sodium-oxygen and
magnesium-aluminum correlations (Ivans et al.\ 1999; Marino et
al.\ 2008; Villanova \& Geisler 2011).
High-resolution VLT/UVES spectra for about one hundred RGB stars
revealed two groups of stars in the Na versus O
plane. Na-rich/O-poor, and Na-poor/O-rich stars populate two different
sequences along the RGB in the $U$ versus $U-B$ CMD. Na-poor stars
populate a spread sequence on the blue side of the RGB, while Na-rich
stars define a narrower red sequence (Marino et al.\ 2008).
M\,4 also features a bimodal HB, which is well populated both on the red
and the blue side of the RR-Lyrae gap. As suggested by Norris (1981)
and Smith \& Norris (1993), HB stars exhibit a bimodal Na and O
distribution similar to what has been found for RGB stars, with red-HB stars
having chemical composition similar to halo field stars and blue-HB
stars being oxygen depleted and sodium enhanced (Marino et al.\ 2011).

While the RGB and HB have been widely investigated from ground-based
photometry and spectroscopy, there are no high-precision photometric
studies on multiple stellar populations among MS and SGB stars.
Recent papers demonstrated that any two-color diagram made with
ultraviolet and visual filters are powerful tools to disentangle
multiple populations and to estimate their helium content (Milone et
al.\ 2012b,c).
The combination of the exquisite visual photometry expected from our
GO-12911 data with photometry in WFC3/UVIS/F275W from GO-12311 (PI: 
Piotto) is expected to provide CMDs with accuracy of a few
milli-magnitudes, therefore allowing us to disentangle the two stellar
components also along the MS of M\,4. 

If we will be able to photometrically distinguish these two stellar
groups among un-evolved stars, not only will we be able to infer the
abundances of CNO and He (as already done for other GCs by Milone et
al.\ 2012b,c), but thanks to our proper motion precision we should
also be able to highlight potential differences in the ensemble
(projected tangential) kinematics of the two stellar groups down to
velocities of a few 100\,m\,s$^{-1}$.
These differences in the kinematics, combined with the radial
distribution, would provide important constraints on the formation and
evolution of the multiple stellar populations (D'Ercole et al.\ 2008; 
2011). 

\subsection{Intermediate-mass central black hole and internal dynamics}
\label{imbh}

\begin{figure}
\includegraphics[width=80mm,height=40mm]{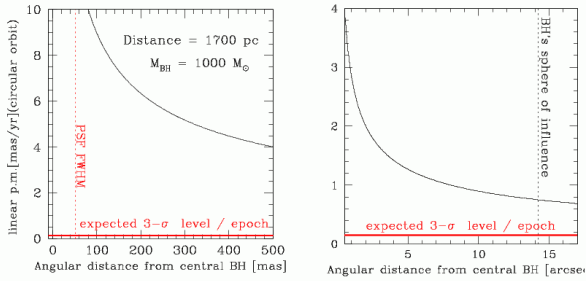}
\caption{Expected size of the proper motion of a MS star around a
  central BH of 1000\,$M_\odot$, as a function of the distance from the BH. The
  two panels show the expected motions at different scales.}
\label{bhs}
\end{figure}

Our investigation will yield accurate relative proper motions for a
large number of stars in the central region of M\,4.  These internal
motions of M\,4 will allow us to search for a possible rise ($v\propto
r^{-0.5}$) in the central velocity dispersion due to the presence of
an intermediate-mass central black hole (IMBH).  Assuming a mass of
$M_{\rm BH}$$=$$10^3M_{\odot}$ (Kormendy \& Richstone 1995) and a
velocity dispersion of $\sigma_0$$\sim$3.5 km/s (Peterson et
al.\ 1995), the black hole would have a sphere of influence of $r_{\rm
  BH}=GM_{\rm BH}/(3\sigma_0^2)\simeq 24,000$ AU, corresponding ---at
the distance of M\,4--- to 14\farcs2, or $\sim$350 WFC3/UVIS pixels.
Moreover, any cluster star particularly close to the central black hole would
have an extremely high velocity (Fig.~\ref{bhs}), which we will be
able to measure trivially (there are about 100 stars within this
radius).  This would be a tremendously exciting finding. 

It is also interesting to consider that if ---by chance--- any
background Bulge star is projected within $\sim
0^{\,\prime\prime}\!\!.\,5$ of the location of the cluster center we might
be able to put even better constraints on the presence of an IMBH
based on lensing constraints.

\subsection{Spectroscopic follow-ups} 

For stars brighter than $V$$\sim$20-21, a spectroscopic follow up is
feasible with 8-10\,m-class telescopes (VLT, SALT, Gemini, Keck,
etc...).  This offers the possibility of obtaining complementary
information which could put strong constraints on many of our
investigations.
In the following we list some of the applications which would benefit
the most from spectroscopic follow-up.
\begin{itemize}
\item 
Detecting the astrometric wobble of the bright component around a dark
companion would provide information on the projected orbit of the
binary system on the plane of the sky.
However, obtaining the RV-curves with a precision of few km s$^{-1}$ 
would not only confirm the wobble detection (being phased with the
astrometric signature) but would also provide the third component
of the motion, and it would enable us to firmly constrain most of the
orbital parameters of the system.  These spectroscopic follow-ups
should be possible for target stars brighter than $V\sim20$-21, with no
neighbor stars within 0.6$^{\prime\prime}$ brighter than $\Delta V =
V_{\rm neighbor}-V_{\rm target} \simeq 2.5$.
\item 
If any detached eclipsing binaries (dEBs) are discovered within
our survey, detailed RV-curve follow-up would allow us to measure
masses, radii and distances of the components (see Sect.~\ref{debs} for
a more extensive discussion).
\item 
Additional RVs to those collected in Sommariva et al.\ (2009) for
stars in M\,4 would enable us not only to obtain the line-of-sight
velocity of the stars in the outskirts of the cluster (which are used
to constrain our dynamical model, Sect.~\ref{dyn_mod}) but also to
better constrain the binary fraction and their orbital parameters
through RV variations.
\item 
If our survey happens to detect an exo-planet candidate transiting
a member MS host star, a detailed RV follow-up study could exclude
obvious false alarms.
This should be possible for stars brighter than $V\sim18$.
\item 
As mentioned in Sect.~\ref{mPOPs}, M\,4 hosts at least two stellar
populations.  Our sub-milli-magnitude photometry and precise
kinematics are expected to clearly disentangle these components even
among stars in the unevolved phases.
We intend to isolate a significant sample of MS stars and collect
high-resolution spectra to study in detail the chemical differences
between the two components down to $V$$\sim$20, as done for
$\omega$\,Cen (Piotto et al.\ 2005) and NGC\,2808 (Bragaglia et
al.\ 2010).
\end{itemize}


\subsection{Photometric binaries} 
\label{pBINs}

The fraction of MS+MS binaries with a relatively high mass-ratio
(i.e.\ made of two luminous components) is also an important
ingredient for the modelling of M\,4. We plan to refine the estimate
of the fraction of MS+MS binaries, and their mass-ratio distribution
using our high precision CMDs and applying the methods and procedures
described in great detail in Milone et al.\ (2012a).

On the basis of our wider color baseline, F467M vs.\ F775W, two times
wider than the F606W vs.\ F814W used in Milone et al.\ (2012a), and of
the anticipated photometric precision, we expect to be able to
estimate the photometric binary distribution down to lower mass-ratios
($q$) than was possible with previous data sets (which were limited to
$q$$>$0.5).
Our dithered images in WFC3/UVIS/F467M and in F775W guarantee
sub-milli-magnitude precision in the study of the MSs of M\,4.

Differential reddening and the presence of multiple stellar
populations within M\,4 will complicate our analyses.
We will follow the method described in Milone et al.\ (2012d) for
the case of NGC\,2808, a cluster with three MSs and comparable
reddening to M\,4.

\subsection{Information from accurate masses and radii of dEBs}
\label{debs}

Despite a number of studies that have presented well ca\-li\-brated
photometry and high-precision CMDs of Galactic GCs, the ages and
helium abundances of their stellar populations remain very uncertain.
This is because of correlated uncertainties in distance, reddening,
color-temperature transformations and metallicity, further complicated
by the interplay between evolutionary times and the unknown helium
contents.  
Detached eclipsing binaries offer the possibility of determining
precise and accurate masses and radii for the system components,
\textit{nearly} independent of model assumptions (Andersen 1991;
Torres et al.\ 2010). If one or both of the binary components
are close to the MS TO, it is possible to put tight
constraints on the age of the system by comparing the position of the
primary and secondary in a mass-radius (${\mathcal M}$-${\mathcal R}$)
diagram to theoretical predictions.  For stellar clusters, such an
analysis has significant advantages: the determination of the masses
and radii is independent of the usual CMD uncertainties such as
distance and reddening (the latter being particularly complex in the
case of M\,4, see Hendricks et al.\ 2012), and one avoids the
difficult process of transforming the effective temperatures and
luminosities of the models to observed colors and magnitudes.

Thus, determining cluster ages in the ${\mathcal M}$-${\mathcal R}$
diagram allows a direct comparison between observations and theory.
We intend to follow-up with 8-10\,m class telescopes the dEBs of
interest that our \hst\/ survey might discover, allowing us to use
similar methods to those of Brogaard et al.\ (2011, 2012): we will
measure masses and radii with accuracies of $\sim$1\% for the dEBs
along with spectroscopic $T_{\rm eff}$ and [Fe/H] values from spectra
of individual components obtained by decomposition of spectra of the
binary stars. This procedure will be possible for the brightest dEBs
(i.e., $V$$<$20).\footnote{
As a demonstration that such suitable dEB systems exist within M\,4,
we note that at the time of the submission of this article, a study of
three dEBs in M\,4 (with $V$$\sim$17) ---all of which are within our
fields--- appeared in the literature (Kaluzny et al.\ 2013). This
study was based on ground-based photometry and on spectroscopic
observations from a 6.5m telescope.
These authors point out that precision time series photometry, which
we will obtain, can reduce their radii uncertainties by $\sim$50\%.
}

The derived values for the individual components of the binaries will
then be used in combination with the observed CMD to tightly constrain
the stellar models and the cluster parameters, including the age and
the helium content, the latter at the $\sim$1\% level.  If a cluster
hosts multiple stellar populations, characterizing at least two dEB
systems for each stellar population would lead to a robust estimate of
age and the He abundance for each sub-population. A difference in He
directly translates into differences in stellar structure, in
particular the radius, leading to a clear separation in 
the ${\mathcal M}$-${\mathcal R}$ plane.  Importantly, it is the
relative locations (not absolute) of these ${\mathcal M}$-${\mathcal R}$ 
sequences that reveal the effect, and, unlike a difference in
age, a difference in He is visible even for stars down the MS.

The combination of geometric distances (see Sect.~\ref{parallax} and
\ref{dist}), CMDs, and $T_{\rm eff}$, radii, and masses of binary
components will provide unprecedented tight constraints on stellar
models of the old populations of the Milky Way.


\subsection{ACS/WFC parallel fields} 
\label{parallels}

Together with our primary observations we will collect a single
ACS/WFC image per orbit, in coordinated-parallel mode.  Unfortunately,
the time needed for the frame-dump buffer does not allow for the
collection of more parallel images.

Located at $\sim$6$'$ from the cluster center, the ACS/WFC parallels 
will map fields well within the cluster tidal radius ($r_{\rm t}$$\sim$30$'$).
The data collected by ACS/WFC will be used for a number of studies in 
the outskirts of the cluster.
As identification of photometric MS+MS binaries does not need temporal
sampling, we will also be able to measure their fractions
(cfr.\,Sect.\,\ref{pBINs}) in the outskirts of M\,4.
Although we will be able to probe only relatively large $q$ in these
parallel fields, it will be possible to study the radial distribution
of these binaries.
Furthermore, since we will see a few of the brightest WDs along the
cooling sequence we will also see WD+MS photometric binaries.  In
addition we shall obtain the MS mass function (as in Bedin et
al.\ 2001), and an empirical measurement of mass segregation, which
will help to tighten the constraints on our dynamical models of M\,4. 
 
ACS/WFC parallel images are significantly deeper, less crowded,
and less absorbed (because of the filters) than the primary WFC3/UVIS observations, and
therefore we also expect them to contain many more back-ground
galaxies at relatively high S/N (see an example of a back-ground
object in the right-panel of Fig.~\ref{acswfc}).
These parallel images will provide absolute high-accuracy proper motions in 
the outskirts of M\,4 (especially if coupled with archival
material, see left and middle panels of Fig.~\ref{obs}).  

Combining the absolute proper motions obtained from the primary WFC3/UVIS
observations in the core, and from the parallel ACS/WFC images in the
outskirts of M\,4, will allow us to measure other fundamental parameters
of the cluster dynamics, such as the anisotropy between radial\footnote{
Here, radial does not mean along the line-of-sight, but the radial direction 
from the cluster center. 
}
and transverse proper motions. 
Anisotropy is an additional constraint on the planned new dynamical
models of M\,4, as described in  
Section~\ref{dyn_mod}. 

Absolute proper motions in the outer fields will enable us to obtain
the average cluster rotation projected on the plane of the sky, as
done by Anderson \& King (2003b, 2004b) for the case of 47\,Tuc.

\begin{figure*}
\begin{center}
\includegraphics[width=152mm,height=152mm]{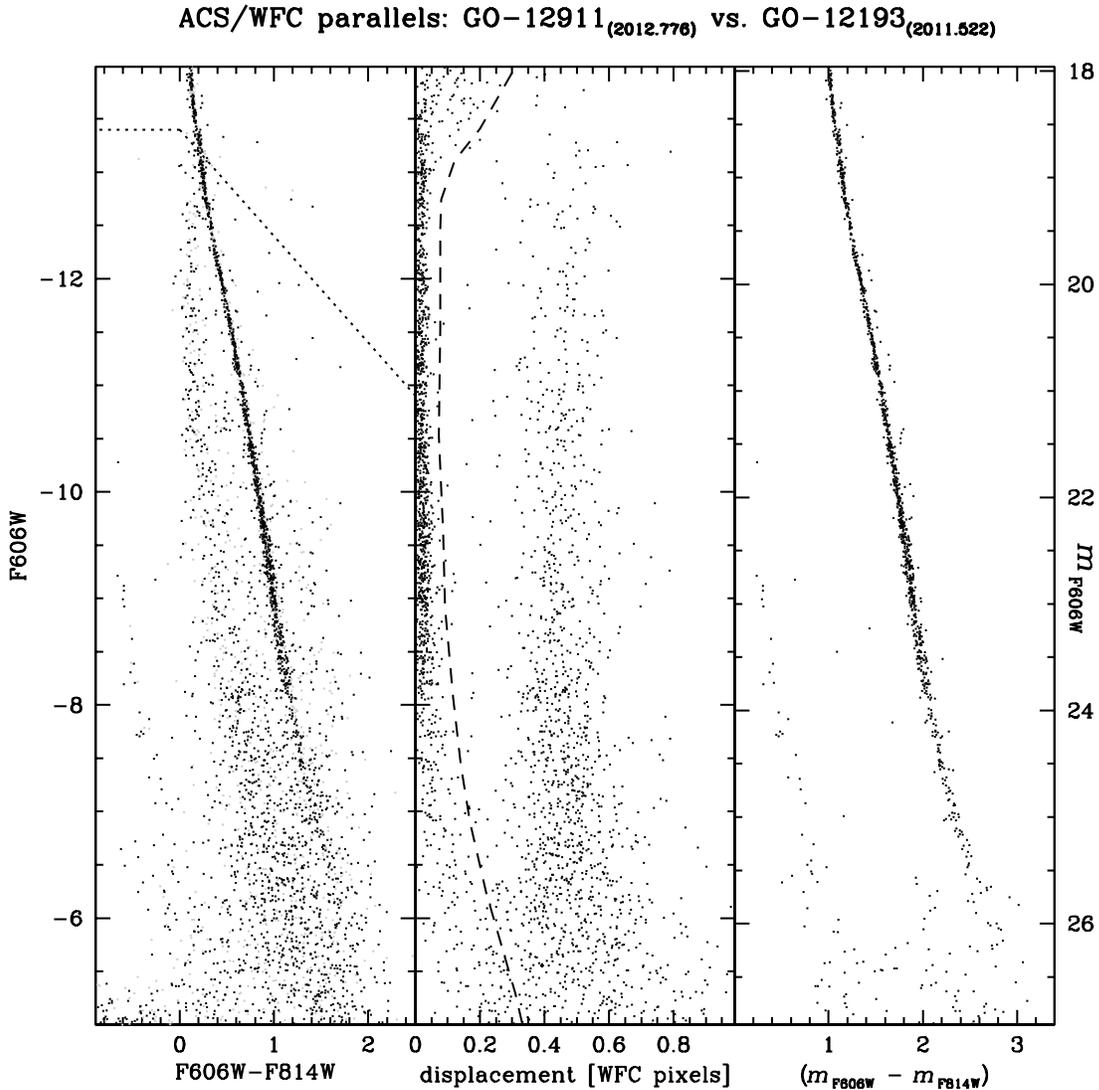}
\caption{
\textit{A parallel field in the outskirts of M\,4:\/}
\textit{(Left:\/)} a portion of the instrumental CMD shown in the left
panel of Fig.~\ref{CMDsACS}.  Stars not measured also within GO-12193, 
or for which it was not possible to obtain a proper motion, are
indicated as grey dots. The dotted line marks the onset of saturation.
\textit{(Center:\/)} the sizes of the proper-motion displacements (in
units of ACS/WFC pixel in $\sim$1.25 years) are plotted vs.\ the
magnitudes. The dashed line marks our proper-motion cutoff criterion
for membership.  Stars on the left of that line are here defined as
members.
\textit{(Right:\/)} shows the calibrated CMD for these member stars. The
main cluster features are considerably decontaminated from field
objects.
}
\label{ACSpms}
\end{center}
\end{figure*}

Figure~\ref{pms} and the related text, presented in Sect.~\ref{pap},
illustrate the proper-motion separation between field and cluster
objects in our M\,4 core field, using the WFC3/UVIS \textit{primary}
observations.
Here, we want to show an example of the proper motion 
precision possible also for objects observed in the ACS/WFC
\textit{parallel} fields, once combined with archival material (when available).
In Fig.~\ref{ACSpms}, we compare positions obtained with our
ACS/WFC/F814W images collected during epoch\,01 (listed in
Table\,\ref{tab01}) with three images collected under program
GO-12193, available from the \hst\,archive.  The latter consist of
just two exposures of 200\,s and 450\,s in ACS/WFC/F606W and one of
400\,s in ACS/WFC/F814W collected in July 9, 2011, for a time
base-line of just $\sim$1.25 years.
In the left panel we show a portion of the CMD presented in
Fig.~\ref{CMDsACS} (in instrumental magnitudes), where stars in common
between GO-12911 and GO-12193 are shown in black, and those not
measured in GO-12193 are shown as gray points. Saturation is indicated
by dotted lines.
The middle panel shows the summed-in-quadrature displacements along
the two axes vs.\ the magnitudes. The units are in ACS/WFC pixels
($\sim$50 mas) in $\sim$1.25 years.  Stars on the left of the dashed
line are assumed to belong to the cluster.
Adopting this selection criterion, on the right panel of
Fig.~\ref{ACSpms} we show the same CMD decontaminated of most 
field objects, and in calibrated magnitudes.


\subsection{Distance to M\,4}
\label{dist}
We mentioned in Sect.~\ref{parallax} our plans to determine the
absolute annual parallax of M\,4 in order to derive the most direct
(and model independent) estimate of the distance of this GC, which is
the closest one to us.
However, we also plan to use three additional methods to obtain
independent estimates of this important parameter of the cluster.
The alternative methods are less direct, but a comparison of the
derived distances with that obtained from the annual parallax offers a
sanity check on the theory adopted to build the cluster dynamical
model and on the adopted stellar evolutionary models.  The methods
are:

\textit{i) Comparison of the cluster internal dispersions of the
  radial velocities $(\sigma_{\rm RV})$ with the dispersion of the
  internal proper motions $(\sigma_{\rm 1-D \mu})$.}
Crudely, combining the linear dimension ($\ell$) and the angular size
($\vartheta$) of the same object (in this case $\sigma_{\rm RV}$ and
$\sigma_{\rm 1-D \mu}$, respectively), provides the distance ($d
\simeq \ell / \vartheta$).  Unfortunately, this estimate is less
straightforward than it looks (Peterson et al.\ 1995).
First it requires an extremely careful evaluation of the error budgets
(measurement uncertainties need to be subtracted in quadrature from the
observed dispersions to get the true intrinsic dispersions). 
Second, the assumption of isotropy of the velocity distribution is not
necessarily justifiable for a real cluster.
This is further complicated by modulations of the dispersion by
unresolved binaries, and by the fact that RVs are available mainly for
bright stars in the outskirts of M\,4, while internal motions are
available mainly for medium-brightness stars in the core of the
cluster.  Hence, in most cases we will not compare the linear and angular
velocities for the same stars. 
The quality of this result will depend sensitively on our dynamical models.


\textit{ii) Double-lined eclipsing binaries.}  
We can hope to detect some of these systems just few magnitudes below the MS TO of M\,4, where
accurate follow-up spectroscopic observations can be made with current
instrumentation at 8-10\,m class telescopes.$^8$ 
%

Indeed, for double-lined eclipsing binaries, the absolute radii of the
components can be determined from observations without any scale
dependence, and this is the basic key to the method. First of all, the
relative radii for each of the two components, measured in units of
the semimajor axis of the orbit, are derived from light-curve analyses
(which also provides the inclination and the orbital period). Second,
the absolute size of the semi-major axis of the orbits is determined
by the semiamplitudes of the radial-velocity variations of the
components through the orbit. If we could angularly resolve the orbit,
the distance to the binary could be established from purely
geometrical methods. However, in the case of two luminous components
at the distance of M\,4, the two photocenters will merge, washing out
most of the astrometric orbital signature.  What we can do, instead,
is to use the relation between surface fluxes and measured fluxes
corrected for interstellar absorption. For the brightest systems,
spectral separation techniques will enable us to separate the
component spectra and perform a spectroscopic analysis to derive
effective temperatures independent of reddening. Combining these with
the measured radii and magnitudes yields the distance (see Brogaard et
al.\ 2011 for a recent application). These methods rely on the
accuracy of the bolometric corrections, and knowledge of the
interstellar extinction, bringing typical internal errors in the
distance of $\sim$5-10\% (see Clausen 2004 for a review).

\textit{iii) Isochrone fitting.} 
The crudest and surely the most model-dependent estimate we can get
of the distance of M\,4 relies on the fit of isochrones to the observed
CMDs. These estimates depend on the absolute calibration of the
photometry, on the knowledge of the reddening law, and on the assumed
chemical compositions for the cluster stars (including CNO and He),
but also, on the input physics of the stellar evolutionary models. It
has to be noted, however, that this method and the previous one are not
independent, since the same stellar model 
(if correct) 
should match
both at the same distance modulus.  Therefore, by analyzing the CMD
and the eclipsing binaries together, a more robust estimate of cluster
parameters, including the distance, can be extracted.

The inter-comparisons between the distance values obtained with these
different methods and that obtained from the annual parallax will tell
us whether or not our dynamical and stellar models are consistent with
observations.
For most clusters, we must rely on distances from stellar models, so
this will be an important cross-check for the method in general.


%

\subsection{Very close binaries revealed through X-ray emission}
\label{xray}

To complement our binary-fraction estimates for those systems with
massive companions and $P$\,$\gtrsim$1 month, we will utilize existing deep
{\it Chandra} observations of M\,4 to search for exotic, closer binary
systems.  Sensitive ($10^{29} < L_\mathrm{X}/[\mathrm{erg~s^{-1}}] <
10^{34}$), high angular resolution ($0.5''$) X-ray observations offer
a very efficient method to locate close binaries in a globular cluster
because nearly every object emitting X-rays at those levels is a close
binary or its progeny.  Combined {\it Chandra} and {\it HST}
observations have shown that these X-ray sources are a heterogeneous
population with a high incidence of compact objects: low-mass X-ray
binaries (LMXBs) in quiescence, cataclysmic variables (CVs), and
millisecond pulsars (MSPs). BY Draconis and RS Canum Venaticorum
magnetically-active binaries (ABs) with main-sequence or (sub-)giant
components are also detected. We refer to Pooley (2010) for a brief
review of GC X-ray studies. The types of binary
interaction that give rise to X-ray emission (e.g.\ mass transfer from
a late-type main-sequence star onto a compact object, or tidal
interaction in ABs) typically take place at relatively short
($\lesssim$1 month) binary periods, so X-ray studies are a nice
accompaniment to the astrometric methods used in our program GO-12911.

While {\it Chandra} is very efficient at finding these close binaries,
classifying them is difficult on the basis of the X-ray data alone, as
often only a handful of photons are detected.  The combination of {\it
  HST} and {\it Chandra} is crucial to determine the types of close
binaries that are found. The first step in source classification is
identifying astrometric matches inside the error circles of {\em
  Chandra} sources; {\em HST\,}'s resolution is critical here.
In many cases we expect to find multiple candidate
counterparts, and the {\em HST} data on M\,4 provide several clues to
finding the true counterpart (if any) among them: proper motion and thus
cluster membership, variability and hence orbital period (see also
Sect.~\ref{var}), or appearance (i.e.\ an extended optical source is
virtually always a background galaxy). Optical colors, which help to
locate a cluster star in the color-magnitude diagram in relation to
the cluster MS, are also important for source
classification. For example, the contribution of an accretion disk or
stream to the optical flux in mass-transferring systems gives rise to
blue colors and often results in strong H$\alpha$ excess emission; ABs
on the other hand often lie above the single-star MS, and
have modest H$\alpha$ excess fluxes.  
The existing ACS/WFC/F658N exposures of M\,4 (GO-10120) in the \hst\/
archive, might turn out to be very useful for this purpose.
X-ray luminosities as derived
from flux and distance estimates and X-ray--to--optical flux ratios
also constrain the binary source class.

Source classification is particularly important when trying to relate
these binary populations to the globular cluster's physical parameters
since different populations have different formation pathways. Pooley
\& Hut (2006) showed that the CV population in a GC is in part formed
through dynamical interactions among the cluster members, but Bassa et
al.\ (2004) demonstrated that the AB population in M\,4 is largely
primordial. On the other hand, van den Berg et al.\ (2013) showed that
both CVs and ABs are under-represented (per unit of cluster mass) in
at least several GCs compared to their numbers in the less
dynamically-active environments of old ($\gtrsim 2$ Gyr) open
clusters; therefore, binary destruction by dynamical interactions may
affect the population of both source classes as well.

These two populations---ABs and CVs---constitute the majority of X-ray
sources in globular clusters, and the existing deep {\it Chandra} data
were taken with the aim of uncovering the bulk of these classes.  {\it
Chandra} observations 7446 and 7447 (PI: Pooley) have a total exposure
time of 100 ksec on M\,4, reaching a luminosity limit of about
$5\times10^{28}~\mathrm{erg~s}^{-1}$.  Based on the X-ray luminosity
function of ABs (Dempsey et al.\ 1997), these X-ray data will find
$\sim$90\% of the ABs in M\,4.  The luminosity function of CVs is less
well-known, so it is difficult to predict what fraction of the CV
population we will discover.  However, very few X-ray observations of
globular clusters reach this luminosity limit, so our combined {\it
Chandra} and {\it HST} observations will give us an excellent observed
population of CVs with which we can constrain the CV X-ray luminosity
function. Comparison of the properties of the AB and CV population in
M\,4 with those of other deeply-studied globular clusters, viz.~the
core-collapsed cluster NGC\,6397 (Cohn et al.\ 2010) and the massive
cluster 47\,Tuc (Heinke et al.\ 2005), will shed light on the effect
of cluster parameters.

We also expect to find more extra-galactic objects among the
M\,4 {\em Chandra} sources, in addition to the one likely
quasi-stellar object found by Bedin et al.\ (2003). This object was
also detected by Bassa et al.\ (2004, 2005) in a study based on
earlier shallower {\em Chandra}\/ data. Such active galaxies can often
be recognized by their blue optical colors, lack of H$\alpha$ excess
emission, and relatively hard X-ray colors.  Extra-galactic reference
objects help to establish the absolute zero point for the proper
motion, see also Sect.~\ref{parallax}.

\subsection{Dynamical modelling} 
\label{dyn_mod}
We intend to update our existing Monte Carlo modelling of the
dynamical evolution of M\,4 (Heggie \& Giersz 2008) for two reasons:
(1) the code has now been much improved, and (2) for the first time we
shall attempt to ensure that the model is consistent with our much
improved knowledge of the binary population, thanks mainly to the
observing program described in this paper.  No $N$-body modelling will
be attempted, as at present this still takes too long; a full-scale
model, based on the initial conditions of Heggie \& Giersz (2008), is
well advanced, but the total run-time will be about 2.5 years.

The Monte Carlo code has already shown its ability and flexibility to
model real star clusters: M67 (Giersz, Heggie \& Hurley 2008), M\,4
(Heggie \& Giersz 2008), NGC6397 (Giersz \& Heggie 2009, Heggie \&
Giersz 2009), and 47\,Tuc (Giersz \& Heggie 2011), and work on M22 is
currently under way.
Nevertheless, the code used in these papers had some drawbacks which
limited its ability to model the dynamics of binaries and the
properties of star cluster exotica (e.g., blue stragglers, millisecond
pulsars, etc).
For example, it is widely assumed that the formation of some blue
stragglers is connected with strong dynamical interactions, which
frequently lead to stellar mergers.  This formation channel could not
be modelled using the cross-sections for binary interactions used in
the old code.
To overcome such limitations, a new version of the code called MOCCA
(MOnte Carlo star Cluster simulAtor) has been developed (Hypki \&
Giersz 2013; Giersz et al.\ 2013).
In addition to the features of the old version, it incorporates the
direct Fewbody integrator (Fregeau et al.\ 2004) for three and
four-body interactions and a new treatment of the escape process
based on Fukushige \& Heggie (2000).
With the addition of the Fewbody integrator the code can follow all
interaction channels that are important for the rate of creation of
various types of objects and binaries observed in star clusters, and
it ensures that energy generation by binaries nearly matches that in an
$N$-body model. The ability of the MOCCA code to follow in detail the
evolution of a population of exotica in globular clusters has been
demonstrated with the example of blue stragglers (Hypki \& Giersz
2013).
Except for some limitations such as spherical symmetry, a Monte Carlo Code
such as MOCCA is at present the most advanced practical code for
simulations of real star clusters.  It can follow the cluster
evolution in detail comparable to an $N$-body code, but it is orders of
magnitude faster.

Our previous modelling of M\,4 was designed to match only the overall
fraction of binaries.  Our new models will be much more ambitious, and
we aim to match not only the very accurate astrometric data obtained in
this program but also the spectroscopic data already collected at
VLT/FLAMES by Sommariva et al.\ (2009, soon to be extended by
Malavolta et al.\ 2013, in preparation) and the photometric binary
observations such as those by Milone et al.\ (2012a, improved with
GO-12911 data).
For the first time, information will be available about the number and
parameters of binaries containing evolved stars (Sect.~\ref{bwmdc}),
photometric (Sect.~\ref{pBINs}), spectroscopic and eclipsing binaries
(Sect.~\ref{var}), ABs and CV's (Sect.~\ref{xray}), and the radial
distribution of some types of binaries (Sect.~\ref{parallels}).
The new features of MOCCA are well suited to this kind of information.
Such data, together with the surface brightness/number density
profile, the luminosity function, and velocity dispersion distribution
in the cluster central region will allow very detailed comparisons
with the results of MOCCA simulations.  The procedure for getting the best
match between the observations and results of simulations
(successfully applied for M\,4, NGC6397, and 47\,Tuc) will be extended
to match observed numbers and properties of different kinds of
binaries and some observed exotica (e.g.\ blue stragglers, CVs) in
M\,4.
Even the new information on the distance
(Sect.~\ref{parallax},~\ref{dist}) and the age (Sect.~\ref{debs})
will be of significant benefit in fitting a model to the observational
data.  In summary, the planned modelling 
will utilize the results 
of several of the projects which are enabled by the new data set.

The new modelling has several underlying aims.  It will allow us to
put more precise constraints than before on the mass, mass ratio,
semi-major axis, period, and eccentricity distributions of primordial
binaries.
It may even be possible to infer whether globular star clusters were
formed with nearly 100\% binary efficiency or with much smaller binary
efficiency.  The number and properties of the binaries with massive
dark components will put very strong constraints on the retention
factor of neutron stars and possibly black holes.  If binaries with a
black-hole component are observed (note that the wobbles should be
largest for them) we may be able to test observationally the fall-back
scenario of the formation of black holes 
(Belczynski, Kalogera \& Bulik 2002). 
The very accurate measurements of the proper motions of a large number
of stars in the cluster core, together with spectroscopic data
(Sommariva et al.\ 2009; Malavolta et al.\ in preparation), will
provide a well-determined velocity dispersion profile, and further
modelling will establish constraints on the possible presence of an
IMBH in M\,4 (Sect.~\ref{imbh}).


\subsection{By-products for  the community}
\label{b-p}

We anticipate that the 600 exposures in F467M and the 120 taken in 
F775W at a variety of dithers and orientations will allow an exquisite average
PSF and distortion solution for these filters.  We will make these products
available to the community.

Breathing-related focus changes cause the PSF in each exposure to differ by
a few percent from the time-averaged PSF.  Since the high-precision measurements
we seek to make in this program are purely differential, we will construct
a tailor-made PSF for each star in each exposure based on its nearest neighbors.
Such fine-scale empirical PSFs should allow an easy study of exactly how
breathing affects the PSF in different locations on the detector.  We will make
these PSF measurements available as well.

The many offsets and orientations of this data set will also allow a direct
empirical evaluation of the quality of CTE restoration in the (forthcoming)
pixel-based algorithm, both in terms of astrometry and photometry.  This should
give the community a direct estimate of how accurate the corrections are for
the typical program, which will not have the redundancy we have here.


To achieve our scientific goals, we will need the most accurate
distortion solution possible.  The data here will allow an exquisite
solution for F467M and will likely improve upon the best-available
solution for F775W (in Bellini et al.\ 2011).


\subsection{Astro-spectro-photometric catalogs and atlases} 
\label{cat}

Upon completion of the observations of GO-12911, we are committed to
release to the astronomical community extensive catalogs and atlases
of the studied regions.  These will include \textit{both} the core of
M\,4 studied with WFC3/UVIS and the ACS/WFC parallel fields in the
outskirts of the cluster.
For objects detected within the studied regions, the catalogs will
include: absolute positions; photometry in the pass-bands of the
program; photometric variability indexes; linear relative proper
motions; parallaxes (when possible); and many quality parameters and
estimates of the errors for each quantity.

Archival \hst\/ material will also be used to improve our motions when
stars are saturated or too faint in the survey's images, or to improve
proper motions when too few multiple observations are available within
GO-12911, or simply to extend the time base-line.  Archival images
will also be used to include photometry in other filters and/or to
improve photometric variability indexes.

We have retrieved and reduced proprietary and archival spectra from
the ESO archive.  At the present time, we have over 10\,000 individual
spectra of M\,4's stars obtained with FLAMES@VLT/ESO, 7\,250 of which
have already been reduced and calibrated (Sommariva et al.\ 2009;
Malavolta et al.\ 2013, in preparation).  These spectra are used to
obtain RVs, atmospheric parameters, and [Fe/H] abundances for many of
the objects in the catalogs (especially in the ACS/WFC parallel
fields).

We aim to release a comprehensive astro-spectro-photometric catalog of
unprecedented depth and accuracy, supported by reference
high-resolution atlases, which will also be made available
electronically.  This effort will result in a significant repository
for the whole astronomical community that can be exploited for a
variety of scientific programmes for many years to come.
In addition to the catalog of average properties, we will also 
release a time-series of flux and position measurements for each star. 

The catalogs themselves will also have many technical applications,
indirectly useful for other scientific topics.
As a simple example, positions and proper motions at a reference
epoch enable us to predict with great accuracy the positions in the
future (or in the past) of the stars in the catalogs.
Therefore, these catalogs provide a ``standard astrometric calibration
field'' which could be used to calibrate the geometric distortion of
many of the present and future instruments, including AO and MCAO
cameras, which generally require high angular resolution and high
spatial density in relatively small fields of view, and which need to
be calibrated at any given epoch.

\begin{table}
\caption{ 
List of planned articles on which our team is working.  The papers
will not necessarily be published in this order, and therefore we
label (as Paper I) only the present work (see text).
}
\label{tab03}
\begin{tabular}{ccc}
\hline
Paper \# & Title               & expected \\ 
\hline
    I   & Overview                           & Feb   2013 \\ 
\dots   & Binary wobbles (main project)      & late  2014 \\ 
\dots   & Variability                        & early 2014 \\ 
\dots   & Exo-planets                        & early 2014 \\ 
\dots   & M\,4 parallax and absolute motions & mid   2014 \\ 
\dots   & Multiple populations within M\,4   & late  2013 \\ 
\dots   & Intermediate-mass Central BH       & mid   2014 \\ 
\dots   & RVs \& Chemistry follow-ups        & ---        \\ 
\dots   & Photometric binaries               & early 2014 \\ 
\dots   & Detached eclipsing binaries        &     2015 \\ 
\dots   & ACS/WFC parallel fields            & mid 2014   \\ 
\dots   & M\,4 distance                      & mid 2014   \\ 
\dots   & X-ray                              & early 2014 \\ 
\dots   & Dynamical model                    & 2016       \\ 
\dots   & Calibration \& by-products         & mid   2014 \\ 
\dots   & Catalog \& atlas                   & mid   2014 \\ 
\hline
\end{tabular}
\end{table}

\acknowledgements

We are grateful to Andrea Dieball for letting us analyze one of the
images in WFC3/UVIS/F390W taken in M\,4 under program GO-12602, while
data were proprietary.
L.R.B., G.P., V.N, S.O., and L.M.\ acknowledge PRIN-INAF 2012 funding
under the project entitled: ``The M4 Core Project with Hubble Space
Telescope''.
J.A., A.B., L.U. and R.M.R.\ acknowledge support from STScI grants GO-12911.
KB Acknowledges support from the Villum Foundation.  A.P.M.\ acknowledges
the financial support from the Australian Research Council through
Discovery Project grant DP120100475. 
V.N.\ and G.P.\ acknowledge partial support by the Universit\`a di Padova
through the ``progetto di Ateneo \#CPDA103591''. 





\begin{thebibliography}{}
%
\bibitem{} Andersen, J.: 1991, A\&ARv~3, 91
\bibitem{} Anderson, J., Bedin, L.\ R.: 2010, PASP~122, 1035
\bibitem{} Anderson, J., King, I.\ R.: 2000, PASP~112, 1360, AK00
\bibitem{} Anderson, J., King, I.\ R.: 2003a, PASP~115,  113, AK03a
\bibitem{} Anderson, J., King, I.\ R.: 2003b, \aj~126, 772
\bibitem{} Anderson, J., King, I.\ R.: 2004a, ACS Instrument Science Report, 2004-15, Baltimore: STScI, AK04a
\bibitem{} Anderson, J., King, I.\ R.: 2004b, \aj~128, 950
\bibitem{} Anderson, J., King, I.\ R.: 2006, ACS Instrument Science Report, 2006-01, Baltimore: STScI, AK06
\bibitem{} Anderson, J., Bedin, L.\ R., Piotto, G., Yadav, R.\ S., Bellini, A.: 2006, A\&A~454, 1029
\bibitem{} Anderson, J., Sarajedini, A., Bedin, L.\ R., et al.: 2008, \aj~135, 2055
\bibitem{} Anderson, J., van der Marel R.: 2010, \apj~710, 1032
\bibitem{} Baggett, S., Anderson, J.: 2012, WFC3 Instrument Science Report, 2012-12, Baltimore: STScI
\bibitem{} Bassa, C., Pooley, D., Lee, H., et al.: 2004, \apj~609, 755
\bibitem{} Bassa, C., Pooley, D., Lee, H., et al.: 2005, \apj~619, 1189
\bibitem{} Bedin, L.\ R., Anderson, J., King, I.\ R., Piotto, G.: 2001, ApJ~560L, 75
\bibitem{} Bedin, L.\ R.; Piotto, G., King, I.\ R., Anderson, J.: 2003, AJ~126, 247
\bibitem{} Bedin, L. R., Cassisi, S., Castelli, F., Piotto, G., Anderson, J., Salaris, M., 
           Momany, Y.,  Pietrinferni, A.: 2005, \mnras~357, 1038
\bibitem{} Bedin, L. R., Piotto, G., Carraro, G., King, I. R., Anderson, J.: 2006, A\&A~460, L27   
\bibitem{} Bedin, L., R., Salaris, M., Piotto, G., Anderson, J., King, I.\ R., Cassisi, S.: 2009, ApJ~697, 965
\bibitem{} Belczynski, K., Kalogera V., Bulik T.: 2002, \apj~572, 407
\bibitem{} Bellini, A., Bedin, L.\ R.: 2009, PASP~121, 1419 
\bibitem{} Bellini, A., Anderson, J., Bedin L.\ R.: 2011, PASP~123, 622 
\bibitem{} Bragaglia, A., Carretta, E., Gratton, R. G. et al.: 2010, ApJ~720, 41
\bibitem{} Brogaard, K., Bruntt, H., Grundahl, F., Clausen, J.\ V., Frandsen, S., 
           Vandenberg, D.\ A., Bedin, L.\ R.: 2011, A\&A~525, 2
\bibitem{} Brogaard, K., VandenBerg, D.\ A., Bruntt, H., et al.: 2012, A\&A~543, 106
\bibitem{} Brown, J.\ A., Wallerstein, G.: 1992, \aj~104, 1818 
%
\bibitem{} Chanam\'e, J., Ramírez, I.: 2012, ApJ~746, 102
\bibitem{} Clausen J.\ V.: 2004, NewAR~48, 679
\bibitem{} Cohn, H.\ N., Lugger, P.\ M., Couch, S.\ M., et al.: 2010, \apj~722, 20
%
\bibitem{} Dempsey, R.\ C., Linsky, J.\ L., Fleming, T.\ A., Schmitt, J.\ H.\ M.\ M.: 1997, \apj~478, 358
\bibitem{} D'Ercole, A., Vesperini, E., D'Antona, F., McMillan, S.\ L.\ W., Recchi, S.: 2008, \mnras~391, 825
\bibitem{} D'Ercole, A., D'Antona, F., Vesperini, E.: 2011, \mnras~415, 1304
\bibitem{} Drake, J.\ J., Smith, V.\ V., Suntzeff, N.\ B.: 1992, \apjl~395, L95 
%
\bibitem{} Ferdman, R.\ D., Richer, H.\ B., Brewer, J., et al.: 2004, AJ~127, 380
\bibitem{} Fregeau, J.\ M., Cheung, P., Portegies-Zwart, S.\ F., Rasio, F.\ A.: 2004, \mnras~352, 1
\bibitem{} Fukushige, T., Heggie, D.\ C.: 2000, \mnras~318, 753
%
\bibitem{} Giersz, M., Heggie, D.\ C.: 2009, \mnras~395, 1173 
\bibitem{} Giersz, M., Heggie, D.\ C.: 2011,  \mnras~411, 2698 
\bibitem{} Giersz, M., Heggie, D.\ C., Hurley, J.\ R.: 2008, \mnras~388, 429
\bibitem{} Giersz, M., Heggie, D.\ C., Hurley J.\ R., Hypki, A.: 2013, MNRAS, in press; arXiv: 1112.6246
\bibitem{} Gilliland, R.\ L.: 2004, ACS Instrument Science Report, 2004-01, Baltimore: STScI
\bibitem{} Gilliland, R.\ L. Brown, T.\ M., Guhathakurta, P., et al.: 2000, \apj~545, L47
\bibitem{} Gratton, R.\ G., Quarta, M.\ L., Ortolani, S.: 1986, A\&A~169, 208 
%
\bibitem{} Harris, W.\ E.: 1996, AJ~112, 1487
\bibitem{} Heggie D.\ C., Hut, P.: 2003, {\sl The Gravitational Million-Body Problem}, Cambridge: Cambridge University Press
\bibitem{} Heggie D.\ C., Giersz, M.: 2008, \mnras~389, 1858  
\bibitem{} Heggie D.\ C., Giersz, M.: 2009, \mnras~397, L46
\bibitem{} Heinke, C.\ O., Grindlay, J.\ E., Edmonds, P.\ D., Cohn, H.\ N., Lugger, P.\ M., 
             Camilo, F., Bogdanov, S., Freire, P.\ C.: 2005, ApJ, 625, 796
\bibitem{} Heintz, W.\ D.:1978, {\em Double Stars}, D.\ Reidel Publishing Company 1978
\bibitem{} Hendricks, B., Stetson, P.\ B., VandenBerg, D.\ A., Dall'Ora, M.: 2012, AJ~144, 25 
\bibitem{} Hills, J.\ G.: 1976, MNRAS~175, 1
\bibitem{} Howard, A.\ W., Marcy, G.\ W., Bryson, S.\ T., et al.: 2012, ApJS~201, 15
\bibitem{} Hut, P., Makino, J.: 1999, Science~283, 501
\bibitem{} Hypki A., Giersz M.: 2013, MNRAS~429, 1221
%
\bibitem{} Ivanova, N.  Heinke, C.\ O., Rasio, F.\ A., Belczynski, K.,
           Fregeau, J.\ M.: 2008, \mnras~386, 553
%
\bibitem{} Ivans, I.\ I., Sneden, C., Kraft, R., P., Suntzeff, N., B., Smith,
           V.\ V., Langer, G.\ E., Fulbright, J. P.: 1999, \aj~118, 1273
%
\bibitem{} Kalirai, J.\ S., Richer, H.\ B., Anderson, J., et al.: 2012, AJ~143, 11
\bibitem{} Kaluzny, J., Thompson, I. B., Rozyczka, M., et al.: 2013, AJ~145, 43
\bibitem{} Kormendy, J., Richstone, D.: 1995, ARA\&A~33, 581
\bibitem{} Kov\'acs, G., Zucker, S., Mazeh, T.: 2002, A\&A~391, 369
\bibitem{} Kroupa, P. 1995: MNRAS~277, 1507
%
\bibitem{} Marino, A.\ F., Villanova, S., Piotto, G., Milone, A.\ P., Momany, Y., 
           Bedin, L.\ R., Medling, A.\ M.: 2008, A\&A~490, 625 
\bibitem{} Marino, A.\ F., Villanova, S., Milone, A.\ P., Piotto, G.,
           Lind, K., Geisler, D., Stetson, P.\ B.: 2011, \apj~730, L16
\bibitem{} Mayor, M., Lovis, C., Pepe, F., S\'egransan, D., Udry, S.: 2011, AN~332, 429
\bibitem{} Meylan, G., Heggie, D.\ C.: 1997, ARAA, 8, 1
\bibitem{} Milone, A.\ P., Villanova, S., Bedin, L.\ R., Piotto, G.,
           Carraro, G., Anderson, J., King, I.\ R., Zaggia, S.: 2006, A\&A~456, 517
\bibitem{} Milone, A.\ P., Piotto, G., Bedin, L.\ R., et al.: 2012a, A\&A~540, A16 
\bibitem{} Milone, A.\ P., Piotto, G., Bedin, L.\ R., et al.: 2012b, \apj~744, 58 
\bibitem{} Milone, A.\ P., Marino, A.\ F., Piotto, G., Bedin, L.\ R., Anderson, J., 
           Aparicio, A., Cassisi, S., Rich, R.\ M.: 2012c, \apj~745, 27
\bibitem{} Milone, A.\ P., Piotto, G., Bedin, L.\ R., Cassisi, S., Anderson, J., 
           Marino, A.\ F., Pietrinferni, A., Aparicio, A.: 2012d, A\&A~537, A77 
%
\bibitem{} Nascimbeni, V., Bedin, L.\ R., Piotto, G., De Marchi, F., Rich, R.\ M.: 2012, A\&A~541, 144	
\bibitem{} Norris, J.\: 1981, \apj~248, 177 
%
\bibitem{} Peterson, R.\ C., Rees, R.\ F., Cudworth, K.\ M.: 1995, \apj~443, 124
\bibitem{} Piotto, G., Villanova, S., Bedin, L.\ R., et al.: 2005, ApJ~621, 777
\bibitem{} Pooley, D., Hut, P.: 2006, \apjl~646, L143
\bibitem{} Pooley, D.: 2010, Proceedings of the National Academy of Science, 107, 7164
%
\bibitem{} Richer, H., Brewer, J., Fahlman, G.\ G., et al.: 2002, ApJ~574, 151   
\bibitem{} Richer, H., Anderson, J., Brewer, J., et al.: 2006, Science~313, 936 
%
\bibitem{} Sarajedini, A., Bedin, L.\ R., Chaboyer, B., et al.: 2007, AJ~133, 1658
\bibitem{} Sigurdsson, S., Richer, H.\ B., Hansen, B.\ M., Stairs, I.\ H., Thorsett, S.\ E.: 2003, Science~301, 5630
\bibitem{} Sippel, A.\ C., Hurley, J.\ R., 2013, MNRAS~430L,30
\bibitem{} Smith, G.\ H., Norris, J.\ E.: 1993, AJ~105, 173 
\bibitem{} Smith, V.\ V., Cunha, K., Ivans, I.\ I., Lattanzio,
           J.\ C., Campbell, S., Hinkle, K.\ H.: 2005, ApJ~633, 392
\bibitem{} Sommariva, V., Piotto, G., Rejkuba, M., Bedin, L.\ R., Heggie, D.\ C.,
           Mathieu, R.\ D., Villanova, S.: 2009, A\&A~493, 947
\bibitem{} Schwarzenberg-Czerny, A.: 1999, ApJ~516, 315	
%
\bibitem{} Torres, G., Andersen, J., Gim\'enez, A.: 2010, A\&ARv~18, 67 
\bibitem{} Trager, S.\ C., King, I.\ R., Djorgovski, S.: 1995, AJ~109, 218 
%
\bibitem{} Ubeda, L., Anderson, J.: 2012, ACS Instrument Science Report, 2012-03, Baltimore: STScI  
\bibitem{} van den Berg, M., Verbunt, F., Tagliaferri, G., Belloni, T., Bedin, L.\ R., 
           Platais, I.: 2013, ApJ, in press; arXiv:1301.2331
%
\bibitem{} van der Marel, R.\ P., Anderson, J., Cox, C., Kozhurina-Platais, V., Lallo, M., 
           Nelan, E.: 2007, ACS Instrument Science Report, 2007-07, Baltimore: STScI
%
\bibitem{} Villanova, S., Geisler, D.: 2011, A\&A~535, A31 
%
\bibitem{} Weldrake, D.\ T.\ F., Sackett, P.\ D., Bridges, T.\ J.: 2008, ApJ~674, 1117 
%
\bibitem{} Zechmeister, M., K\"urster, M.: 2009, A\&A~496, 577	
%
\bibitem{} Zhou, J.,  Xie, J., Liu, H., Zhang, H., Sun, Y.: 2012, RAA~12, 1081
%
\end{thebibliography}
\end{document}